\documentclass[prd,twocolumn,nofootinbib,showpacs]{revtex4-1}

\usepackage{amsfonts}
\usepackage{amsmath}
\usepackage{amssymb}
\usepackage{bm}
\usepackage{dcolumn}
\usepackage{graphicx}
\usepackage{graphics}
\usepackage[latin1]{inputenc}
\usepackage{latexsym}
\usepackage{rotating}
\usepackage[colorlinks=true]{hyperref}
\usepackage{xspace} 
\usepackage[usenames]{color}
\usepackage{mathrsfs}

\usepackage{ulem}
\definecolor {darkgreen}{rgb}{0.2,0.7,0.2}

\newcommand\be{\begin{equation}}
\newcommand\ba{\begin{eqnarray}}
\newcommand\ee{\end{equation}}
\newcommand\ea{\end{eqnarray}}

\newcommand\bw{\begin{widetext}}
\newcommand\ew{\end{widetext}}

\newcommand{\nn}{\nonumber}

\newcommand{\N}{{\mbox{\tiny N}}}
\newcommand{\firstPN}{{\mbox{\tiny 1PN}}}
\newcommand{\BH}{{\mbox{\tiny BH}}}

\newcommand{\PP}{{\mbox{\tiny PP}}}
\newcommand{\FS}{{\mbox{\tiny FS}}}

\newcommand{\stat}{{\mbox{\tiny stat}}}

\newcommand{\lambdabarnew}{\bar{\lambda}}
\newcommand{\sigmabar}{\bar{\sigma}}
\newcommand{\etabar}{\bar{\eta}}

\newcommand{\mrm}{\mathrm}

\begin{document}
\title{Multipole Love Relations}

\author{Kent Yagi}
\email{kyagi@physics.montana.edu}
\affiliation{Department of Physics, Montana State University, Bozeman, MT 59717, USA.}


\date{\today}

\begin{abstract} 

Gravitational-wave observations in the near future may allow us to measure tidal deformabilities of neutron stars, which leads us to the understanding of physics at nuclear density. In principle, the gravitational waveform depends on various tidal parameters, which correlate strongly. Therefore, it would be useful if one can express such tidal parameters with a single parameter.
Here, we report on universal relations among various $\ell$-th (dimensionless) electric, magnetic and shape tidal deformabilities in neutron stars and quark stars that do not depend sensitively on the equation of state. 
Such relations allow us to break the degeneracy among the tidal parameters. In this paper, we focus on gravitational waves from non-spinning neutron-star binary inspirals. We first derive the leading contribution of the $\ell$-th electric and $\ell=2$ magnetic tidal deformabilities to the gravitational-wave phase, which enters at $2\ell +1$ and $6$ post-Newtonian orders relative to the leading Newtonian one respectively. We then calculate the useful number of gravitational-wave cycles and show that not only the $\ell=2$ but also $\ell=3$ electric tidal deformabilities are important for parameter estimation with third-generation gravitational-wave detectors such as LIGO III and Einstein Telescope. 
Although the correlation between the $\ell=2$ and $\ell=3$ electric tidal deformabilities deteriorate the measurement accuracy of the former deformability parameter, one can increase its measurement accuracy significantly by using the universal relation. We provide a fitting formula for the LIGO III noise curve in the Appendix.

\end{abstract}



\maketitle

\section{Introduction}

\subsection*{Background}

The history of Newtonian tides goes back to 1911. The theory of Newtonian tides was established by Love~\cite{love} who introduced two tidal parameters, the first and second apsidal constants (or tidal Love numbers). The former measures a dimensionless ratio between the $\ell$-th multipolar deformation of a star and the $\ell$-th multipolar external tidal potential, while the latter refers to a dimensionless ratio between the $\ell$-th induced multipole moment and the the $\ell$-th multipolar external tidal potential. 

A relativistic theory of Love numbers was formalized by two independent groups~\cite{damour-nagar,binnington-poisson}. In the relativistic context, three types of Love numbers exist, the shape, electric and magnetic. The shape and electric tidal Love numbers correspond to the first and second Love numbers in the Newtonian limit respectively, while Newtonian analogue does not exist for the magnetic tidal Love numbers. The authors of Ref.~\cite{damour-nagar,binnington-poisson} restricted themselves to nondynamical tides, while the formalism has been extended to include dynamic tides in~\cite{ferrari,maselli-affine,chakrabarti1,chakrabarti2} using either the post-Newtonian (PN) affine framework~\cite{ferrari-affine1,ferrari-affine2} or effective field theory approach~\cite{goldberger1,goldberger2}. 

Tidal Love numbers, or tidal deformabilities to be more precise, enter in the phase of gravitational waves (GWs) from neutron-star (NS) binary inspirals, and hence future observations may reveal fundamental properties in nuclear physics. Such inspirals are one of the most promising sources for second-generation GW interferometers such as Adv.~LIGO~\cite{ligo}, Adv.~VIRGO~\cite{virgo} and KAGRA~\cite{kagra}. The leading finite-size effect in the GW phase enters at 5PN order relative to the Newtonian one and comes from the $\ell=2$ electric tidal deformability. The $\ell=2$ electric tidal Love number of NSs was first calculated by Hinderer~\cite{hinderer-love}, whereas the one for quark stars (QSs) was calculated e.g. in~\cite{hinderer-lackey-lang-read,postnikov}. Flanagan and Hinderer~\cite{flanagan-hinderer-love} showed that the tidal deformability can be measured by Adv.~LIGO, and much work followed~\cite{read-love,hinderer-lackey-lang-read,kyutoku,kiuchi,hotokezaka,vines2,lackey,damour-nagar-villain,baiotti,pannarale,bernuzzi,hotokezaka2,lackey-kyutoku-spin-BHNS,read-matter,delpozzo,maselli-GW-EOS}. The systematic errors on tidal parameters with GW observations due to mis-modeling of the waveform template are discussed in~\cite{favata-sys,kent-sys}. The effect of higher order electric tidal deformabilities in GW phase is discussed in~\cite{flanagan-hinderer-love,hinderer-lackey-lang-read,damour-nagar-villain}, while the effect of the $\ell=2$ magnetic tidal deformability on NS stability was calculated in~\cite{favata}. The shape Love number was used to investigate the possibility of resonant shattering of the NS crust~\cite{tsang}. Specifically, the authors in~\cite{damour-nagar-villain} recommend to include such higher order contribution into model parameters when carrying out GW data analysis and give a possible way to decrease the number of tidal parameters. 
We investigate this possibility further by using the universal relations among the tidal deformabilities that do not depend strongly on the NS internal structures.

Universal relations among the NS and QS observables have been investigated by many authors. Universal relations among NS oscillation modes have been investigated deeply in~\cite{andersson-kokkotas-1996,andersson-kokkotas-1998,benhar1999,benhar2004,tsui-leung,lau}. Bejger and Haensel~\cite{bejger} and Lattimer and Schutz~\cite{lattimer-schutz} found a universal relation between the rescaled moment of inertia and the compactness. Such relation helps in measuring the moment of inertia using double pulsar binary J0737-3039~\cite{burgay,lyne,kramer-double-pulsar,kramer-wex}. Extending these, Urbanec \textit{et al}.~\cite{urbanec} found a universal relation between the rescaled quadrupole moment and the compactness. Several authors report on the EoS-independent relation between the cutoff frequency of GWs from NS binaries and compactness~\cite{kyutoku,kiuchi}.

Recently, the author and Yunes~\cite{I-Love-Q-Science,I-Love-Q-PRD} found yet another relations between the (dimensionless) $\ell=2$ electric tidal deformability, moment of inertia and quadrupole moment of NSs and QSs that do not depend sensitively on the internal structure of the stars. Such I-Love-Q relations have several applications. On the astrophysical front, one measurement of the I-Love-Q trio would automatically determine the other two. On the GW physics front, the universal relations help to break the degeneracy between spins and other parameters. On the fundamental physics front, the independent measurement of any two of the trio allows us to perform a model-independent and equation of state (EoS)-independent test of general relativity (GR). 

Such universal relations have been confirmed by follow-up studies such as~\cite{lattimer-lim,maselli,I-Love-Q-B,doneva-rapid}. Maselli \textit{et al}.~\cite{maselli} adopted the PN affine framework to study the relations under dynamical tidal perturbations and found that the universality is still preserved to within 10\% accuracy for NS binary inspirals with GW frequencies $f \lesssim 900$ Hz. Haskell \textit{et al}.~\cite{I-Love-Q-B} investigated the effect of magnetic fields and found that the universality holds unless the star has rather large magnetic fields ($\gtrsim 10^{12}$ G) and a rather long period ($\gtrsim 10$ s). Doneva \textit{et al}.~\cite{doneva-rapid} looked at the I-Q relations for rapidly-rotating NSs and QSs. They found that the relations deviate away from those of slowly-rotating NSs and QSs, but if one fixes the NS spin frequency, the relations still only depend weakly on the modern realistic EoS, as qualitatively expected and discussed in~\cite{I-Love-Q-PRD}. They also found that the EoS-dependence becomes larger as one increases the NS spin frequeny. They confirmed that one can still apply the universal relations found for slowly-rotating NSs if they spin slower than a few hundred Hz. Baubock \textit{et al}.~\cite{baubock} used the I-Q relation, together with universal relations among the NS  spin angular momentum, ellipticity and compactness, and showed that such relations help to extract parameters from the NS emission lines.

\begin{figure}[htb]
\begin{center}
\includegraphics[width=8.5cm,clip=true]{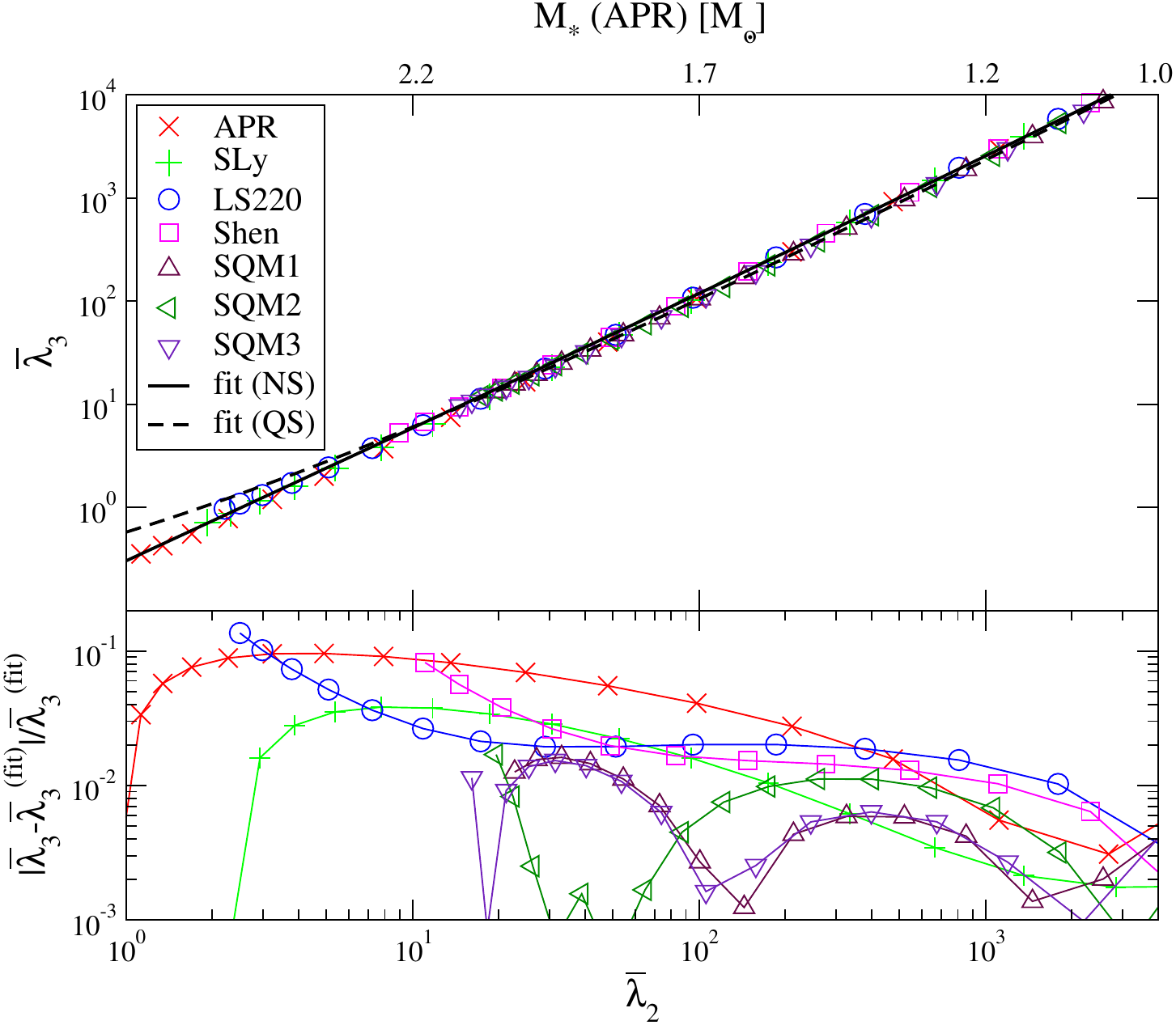}  
\caption{\label{fig:lambdabar3-lambdabar2} 
(Color online) (Top) Universal relation between the $\ell=2$ and $\ell=3$ dimensionless electric tidal deformabilities for NSs (APR, SLy, LS220 and Shen EoSs) and QSs (three SQM EoSs). The single parameter along the curves is the mass or compactness. We construct fitting formulas for the realistic EoSs given by Eq.~\eqref{fit}, which are shown with black solid (NS) and dashed (QS) curves. As a reference, we show the NS mass for the APR EoS on the top axis. Observe how the relation with each EoS lie on top of each other. 
(Bottom) Fractional difference of each curve to the fitting formulas. 
}
\end{center}
\end{figure}

\subsection*{Executive Summary}

Here, we give a brief summary of our results. First, following~\cite{damour-nagar}, we calculate the $\ell=2$, 3, 4 electric, $\ell=2$ magnetic and $\ell=2$, 3, 4 shape tidal deformabilities of NSs and QSs. 
We then show new universal relations for NSs and QSs, similar to the I-Love-Q ones, among such tidal deformabilities.
The relation between the $\ell=2$ and $\ell=3$ dimensionless electric tidal deformabilities ($\lambdabarnew_\ell$) is shown in the top panel of Fig.~\ref{fig:lambdabar3-lambdabar2} for various EoSs. Similar relations hold among $\lambdabarnew_\ell$ and the $\ell=2$ magnetic tidal deformability. The single parameter along the curves is the mass or compactness. Notice how each curve lies on top of each other. In the bottom panel, we show the fractional difference between numerical and fitted values. One sees that the universality holds within $\mathcal{O}(10)$\% accuracy, which is not as universal as the I-Love-Q relations~\cite{I-Love-Q-Science,I-Love-Q-PRD}, where the latter hold to $\mathcal{O}(1)$\%. We also find universal relations among the $\ell=2$, 3 and 4 shape deformabilities. 

Such new relations have similar applications to the I-Love-Q ones~\cite{I-Love-Q-Science,I-Love-Q-PRD}. Especially, such relations help to reduce the number of finite-size effect parameters in the GW phase for a binary NS inspiral. In fact, the relations suggest that one can express finite-size effects with a single parameter, the $\ell=2$ electric tidal deformability. We derive that the leading contribution of the $\ell$-th $(\ell \geq 2)$ electric and magnetic tidal deformabilities to the GW phase enters at $2 \ell + 1$ and $3 \ell$ PN orders, respectively, relative to the leading Newtonian one. We then calculate the useful number of GW cycles~\cite{damour-useful} and show that both $\ell=2$ and $\ell=3$ electric tidal deformabilities may affect GW parameter estimation for third-generation GW detectors such as LIGO III~\cite{LIGO3-noise} and Einstein Telescope (ET)~\cite{et,ET-noise}.

\begin{figure*}[thb]
\begin{center}
\includegraphics[width=8.5cm,clip=true]{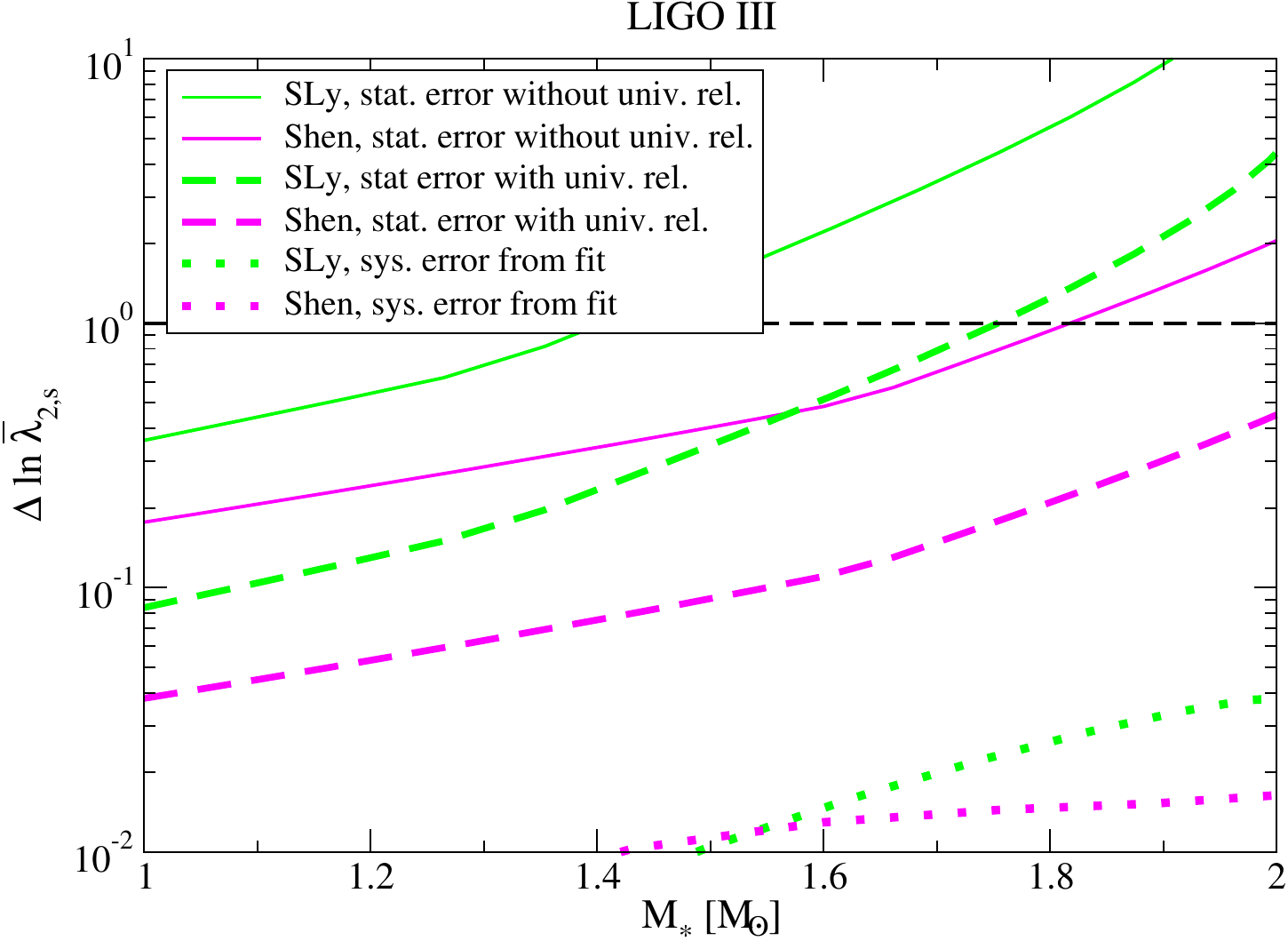}  
\includegraphics[width=8.5cm,clip=true]{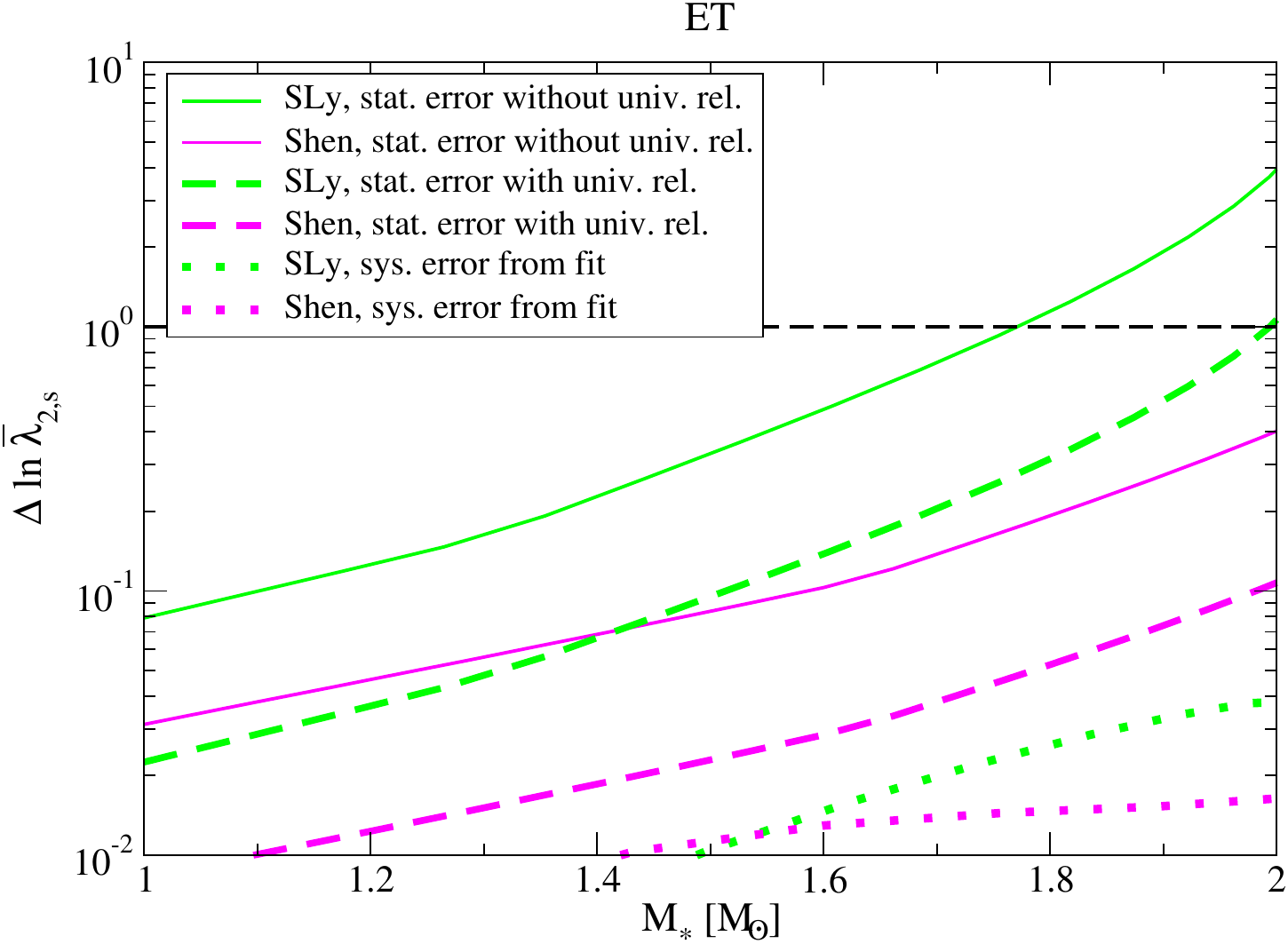}  
\caption{\label{fig:delta-lambda2-LIGO3} 
(Color online) Statistical errors on the averaged $\ell=2$ dimensionless electric tidal deformability $\lambdabarnew_{2,s}$ against a NS mass with and without using the universal relation using LIGO-III~\cite{LIGO3-noise} (left) and ET~\cite{et,ET-noise} (right), and systematic errors due to the fitting (shown in the bottom left panel of Fig.~\ref{fig:lambdabar3-lambdabar2}) with the SLy and Shen EoSs. We assume that one detects a GW signal from a non-spinning, equal-mass NS/NS binary at 100Mpc. $\lambdabarnew_{2,s}$ reduces to $\lambdabarnew_{2}$ of each body for an equal-mass binary. By using the relation, one can remove the $\ell=3$ averaged dimensionless electric tidal deformability $\lambdabarnew_{3,s}$ from the model tidal parameters. Observe that one can reduce statistical errors by a factor of 5 if one uses the universal relation. Notice also that the systematic errors due to the fitting are much smaller than the statistical errors, which validates the use of the fitting function of the universal relation to reduce the number of parameters. 
}
\end{center}
\end{figure*}

We further investigate the effect of the universal relations on the GW physics. We calculate statistical errors on the averaged (among the two binary constituents) $\ell=2$ electric tidal deformability $\lambdabarnew_{2,s}$ using a Fisher analysis~\cite{cutlerflanagan}, assuming that we detect GW signals from a non-spinning NS binary inspirals. For an equal-mass binary, $\lambdabarnew_{2,s}$ is identical to $\lambdabarnew_{2}$ of each body.

Figure~\ref{fig:delta-lambda2-LIGO3} presents the statistical errors on $\lambdabarnew_{2,s}$ against each NS mass for an equal-mass NS/NS binary with LIGO III (left) and ET (right) with the SLy and Shen EoSs. We take both $\lambdabarnew_{2,s}$ and $\lambdabarnew_{3,s}$ into account, and show the statistical errors with and without using the universal $\lambdabarnew_{3,s}$--$\lambdabarnew_{2,s}$ relation (which is the same as the $\lambdabarnew_{3}$--$\lambdabarnew_{2}$ relation shown in Fig.~\ref{fig:lambdabar3-lambdabar2}). 
If one does not use the universal relation, one has a large statistical error due to a strong correlation between these two tidal parameters. However, once one uses the universal relation, one can reduce the statistical error to the level where one only includes $\lambdabarnew_{2,s}$. This is because the relation breaks the degeneracy between $\lambdabarnew_{2,s}$ and $\lambdabarnew_{3,s}$ as such relation allows us to express $\lambdabarnew_{3,s}$ in terms of $\lambdabarnew_{2,s}$. 
One sees that one can reduce the statistical error by a factor of 5 if one uses the universal relation.
One also sees that the systematic errors due to the uncertainty in the EoS of the universal relation shown in the bottom panel of Fig.~\ref{fig:lambdabar3-lambdabar2} are smaller than the statistical errors as shown in Fig.~\ref{fig:delta-lambda2-LIGO3}, which validates our use of universal relation.

\subsection*{Organization and Conventions}

The organization of this paper is as follows. In Sec.~\ref{sec:Love0}, we define three different types of tidal Love numbers and deformabilities. In Sec.~\ref{sec:Love}, we follow~\cite{damour-nagar} and calculate such tidal parameters for NSs and QSs, by first constructing the spherically-symmetric NSs and QSs, and then consider small tidal perturbations. In Sec.~\ref{sec:universal}, we show universal relations, give fitting functions and perform an analytic analysis in the Newtonian limit. In Sec.~\ref{sec:GWs}, we explain how such relations bring about benefits to GW physics. We first show how the tidal parameters would enter in the GW phase of the compact binary inspiral. We then calculate the useful number of GW cycles of each finite-size terms in the GW phase for Adv.~LIGO, LIGO III and ET. Furthermore, we perform a Fisher analysis and estimate statistical errors on the tidal parameter with and without using the universal relation. We conclude in Sec.~\ref{sec:conclusions} and explain possible avenues for future work. In App.~\ref{app:phase2}, we show a derivation of the finite-size effect phase terms due to the $\ell$-th electric and $\ell=2$ magnetic tidal deformabilities. In App.~\ref{app:phase}, we explicitly show some of the point-particle and finite-size effect terms in the GW phase.  In App.~\ref{app:LIGOIII}, we construct a fitting formula for the LIGO III noise curve.

All throughout the paper, we follow mostly the conventions of Misner, Thorne and Wheeler~\cite{MTW}. We use the Greek letters $(\alpha, \beta, \cdots)$ to denote spacetime indices, while we use the Roman letters $(a,b,\cdots)$ to denote the space indices. The metric, and especially the flat one, are denoted by $g_{\mu \nu}$ and $\eta_{\mu\nu}$ respectively, and they have signatures $(-,+,+,+)$. We use the notation $A_{L} \equiv A_{\langle a_1 a_2 \cdots a_\ell \rangle}$ to denote symmetric trace-free (STF) tensors~\cite{thorne-MM} with $\ell$ indices. We use the geometric units of $G=1=c$.

\section{Tidal Love Numbers and Deformabilities}
\label{sec:Love0}

In this section, we define various kinds of relativistic tidal Love numbers and deformabilities of non-rotating bodies~\cite{damour-nagar,binnington-poisson}, which play the central role in this paper. We follow~\cite{damour-nagar} and work in the spherical coordinates, while Ref.~\cite{binnington-poisson} works in the light-cone coordinates\footnote{Ref.~\cite{binnington-poisson} also shows that the tidal deformabilities are gauge invariant.}. 

\subsection{Electric and Magnetic}

First, let us define the electric and magnetic tidal parameters. One can calculate the effect of internal structures of bodies on their motion and radiation from the system by solving the outer and inner problems and perform matching. The former requires solving the field equations in which the bodies are characterized by multipole moments, and one adopts a global weak-field expansion of the metric as $g_{\mu \nu} (x) = \eta_{\mu\nu} + h_{\mu\nu}(x)$ with the global coordinates $x^{\mu}$. On the other hand, the latter takes into account the effect of other bodies on each body~\cite{damour-effacement} with the metric expanded as $G^A_{\mu \nu}(X_A) = G^{A,(0)}_{\mu \nu} (X_A) + H^A_{\mu\nu}(X_A)$. Here, $G^{A,(0)}_{\mu \nu} (X_A)$ is the metric for an isolated body $A$ in local inner coordinates $X_A^\mu$ while $H^A_{\mu\nu}(X_A)$  refers to a perturbation due to other bodies.
We restrict ourselves to stationary fields throughout this paper. 

The local gravito-electric and gravito-magnetic fields $E_a^A$ and $B_a^A$ are defined by~\cite{DSX1,DSX2,DSX3,DSX4}
\be
E_a^A = \partial_a W^A, \quad B_a^A = -4 \epsilon_{abc} \partial_b W_c^A\,,  
\ee
where the potentials $W^A$ and $W_a^A$ are defined through the metric $G_{\mu\nu}^A$ via
\be
G_{00}^A = -\exp \left( -2 W^A \right), \quad G_{0a}^A = -4 W_a^A\,.
\ee
From the externally-generated parts of the fields $\bar{E}_a^A$ and $\bar{B}_a^A$, which are proportional to $R^{\ell -1}$ for $R \equiv |X^a| \rightarrow \infty$ in $E_a^A$ and $B_a^A$, one can define the $\ell$-th gravito-electric and gravito-magnetic relativistic tidal moments $G_L^A$ and $H_L^A$ as\footnote{For a tidal perturbation around a flat spacetime, tidal moments are related to the Weyl tensor~\cite{binnington-poisson}.} 
\ba
G_L^A & \equiv & \partial_{\langle L-1} \bar{E}^A_{a_\ell \rangle}|_{X^a \rightarrow 0}\,, \\
H_L^A & \equiv & \partial_{\langle L-1} \bar{B}^A_{a_\ell \rangle}|_{X^a \rightarrow 0}\,.
\ea

Next, we define the internally-generated mass and current multipole moments of the $A$-th body, $M_L^A$ and $S_L^A$, which scale as $R^{- (\ell +1)}$ at spatial infinity. Such multipole moments are defined through the internally-generated part of the potentials $W^{+ A}$ and $W^{+A}_a$ via
\ba
W^{+A} (X) &=& \sum_{\ell = 0} \frac{(-1)^\ell}{\ell !} \partial_{L} \left( \frac{M_L^A}{R} \right)\,, \\
W^{+A}_a (X) &=& \sum_{\ell = 1} \frac{(-1)^\ell \ell}{(\ell +1) \ell !} \epsilon_{abc} \partial_{b L-1} \left( \frac{S^A_{c L-1}}{R} \right)\,, \\
\ea
modulo a gauge transformation~\cite{damour-nagar}. Notice that for non-rotating bodies, $M_L^A$ and $S_L^A$ correspond to tidally-induced multipole moments.

Having the above quantities at hand, we define the $\ell$-th electric and magnetic tidal deformabilities $\lambda_\ell$ and $\sigma_\ell$ as\footnote{In~\cite{I-Love-Q-PRD,flanagan-hinderer-love}, $\lambda_\ell$ is called the $\ell$-th electric tidal Love number.}~\cite{DSX2} 
\be
\label{eq:linear-tidal}
M_L^A = \lambda_\ell^A G_L^A, \quad S_L^A = \sigma_\ell^A H_L^A\,.
\ee
We also define the \emph{dimensionless} $\ell$-th electric and magnetic tidal deformabilities $\lambdabarnew_\ell$ and $\sigmabar_\ell$ by appropriately normalizing $\lambda_\ell$ and $\sigma_\ell$ by the mass of a star $M_*$ as
\be
\lambdabarnew_\ell \equiv \frac{\lambda_\ell}{M_*^{2\ell +1}}, \quad \sigmabar_\ell \equiv \frac{\sigma_\ell}{M_*^{2\ell +1}}\,.
\ee
One can define the $\ell$-th relativistic electric and magnetic tidal Love numbers $k_\ell$ and $j_\ell$ as~\cite{damour-nagar}
\ba
k_\ell &\equiv & \frac{(2 \ell -1)!!}{2} C^{2 \ell +1} \lambdabarnew_\ell\,, \\
 j_\ell &\equiv & \frac{4 (\ell +2) (2 \ell -1)!!}{\ell -1} C^{2 \ell +1} \sigmabar_\ell\,,
\ea
where $C$ is the compactness of the star defined by $C \equiv M_* / R_*$ with $R_*$ denoting the radius of the star. $k_\ell$ corresponds to the second apsidal constant in the Newtonian limit while there is no analogue for $j_\ell$ in such limit.

\subsection{Shape}

Next, we move onto defining the shape tidal parameters~\cite{damour-nagar}. First, we decompose the external disturbing potential $U(r,\theta)$ as
\be
U(r,\theta) = \sum_{\ell} U_\ell(r) P_\ell (\cos\theta)\,,
\ee
where $P_\ell(x)$ denotes a Legendre polynomial. Then, one can define the shape Love number $h_\ell$ as~\cite{damour-nagar}
\be
\label{eq:shape}
\frac{\delta R_\ell}{R_*}  = h_\ell \frac{U_\ell (R_*)}{C}\,,
\ee
where $\delta R_\ell/R_*$ denotes the $\ell$-th fractional deformation of the surface of the star. Similar to the relation between $\lambdabarnew_\ell$ and $k_\ell$, we define the dimensionless shape tidal deformability $\etabar_\ell$ as
\be
\etabar_\ell \equiv  \frac{2}{(2 \ell -1)!!} \frac{1}{C^{2 \ell +1}} h_\ell\,. 
\ee
$h_\ell$ corresponds to the first apsidal constant in the Newtonian limit.

\section{Tidally-Deformed Compact Stars and Tidal Deformabilities} 
\label{sec:Love} 

In this section, we explain how one can construct a tidally-deformed NS solution\footnote{A QS solution can be obtained in a similar manner. The only difference is the choice of the EoS.} to calculate the tidal deformabilities. We explain the construction of an isolated, non-spinning NS as a background in Sec.~\ref{sec:bg} and then perturb this solution to obtain various tidal deformabilities in Sec.~\ref{sec:pert}. We follow~\cite{damour-nagar} and calculate the $\ell=2$, 3, 4 electric, $\ell=2$ magnetic and $\ell=2$, 3, 4 shape dimensionless tidal deformabilities.

\subsection{Background}
\label{sec:bg}

We begin by constructing a spherically symmetric, isolated non-spinning NS solution. We impose the metric ansatz as~\cite{hartle1967}
\be
ds_0^2 = -e^{\nu(r)} dt^2 + e^{\lambda (r)} dr^2 + r^2 (d\theta^2 + \sin^2 \theta d\varphi^2)\,.
\ee
The matter stress-energy tensor $T_{\mu \nu}$ is given by
\be
T_{\mu \nu} = (\rho + p) u_{\mu} u_{\nu} + p g_{\mu\nu}\,,  
\ee
where $p$ and $\rho$ are the NS pressure and energy density, respectively, and $u^\mu$ is the fluid four-velocity given by
\be
u^{\mu} = (e^{-\nu/2}, 0, 0,0)\,.
\ee
Here, $u^0$ is obtained from the normalization condition $u_\mu u^\mu = -1$. One substitute the above ansatz to the field equations to yield the Tolman-Oppenheimer-Volkoff equations:
\ba
\label{tt-zeroth}
\frac{d M}{dr} &=& 4 \pi r^2 \rho\,, \\
\label{rr-zeroth}
\frac{d \nu}{dr} &=& 2\frac{4 \pi r^3 p + M}{r(r-2M)}\,, \\
\label{TOV-zeroth}
\frac{dp}{dr} &=& -\frac{(4\pi r^3 p + M) (\rho + p)}{r(r-2M)}\,,
\ea
where $M(r)$ is defined by
\be
M(r) \equiv \frac{\left[ 1- e^{-\lambda (r)} \right] r}{2}\,.  
\ee
Outside a star, $M(r)$ becomes a constant $M_*$, which corresponds to the NS ADM mass.
One needs to solve Eqs.~\eqref{tt-zeroth}--\eqref{TOV-zeroth}, together with a barotropic EoS, $p=p(\rho)$. The interior solution is obtained by imposing regularity at the NS center with a choice of central density $\rho_c$. The NS surface is defined to be where the pressure vanishes, i.e. $p(R_*) = 0$ where $R_*$ is the NS radius.  The exterior solution is obtained by imposing asymptotic flatness at spatial infinity. The integration constants are then determined by matching the interior and exterior solutions at the NS surface. 

\begin{figure}[thb]
\begin{center}
\includegraphics[width=8.5cm,clip=true]{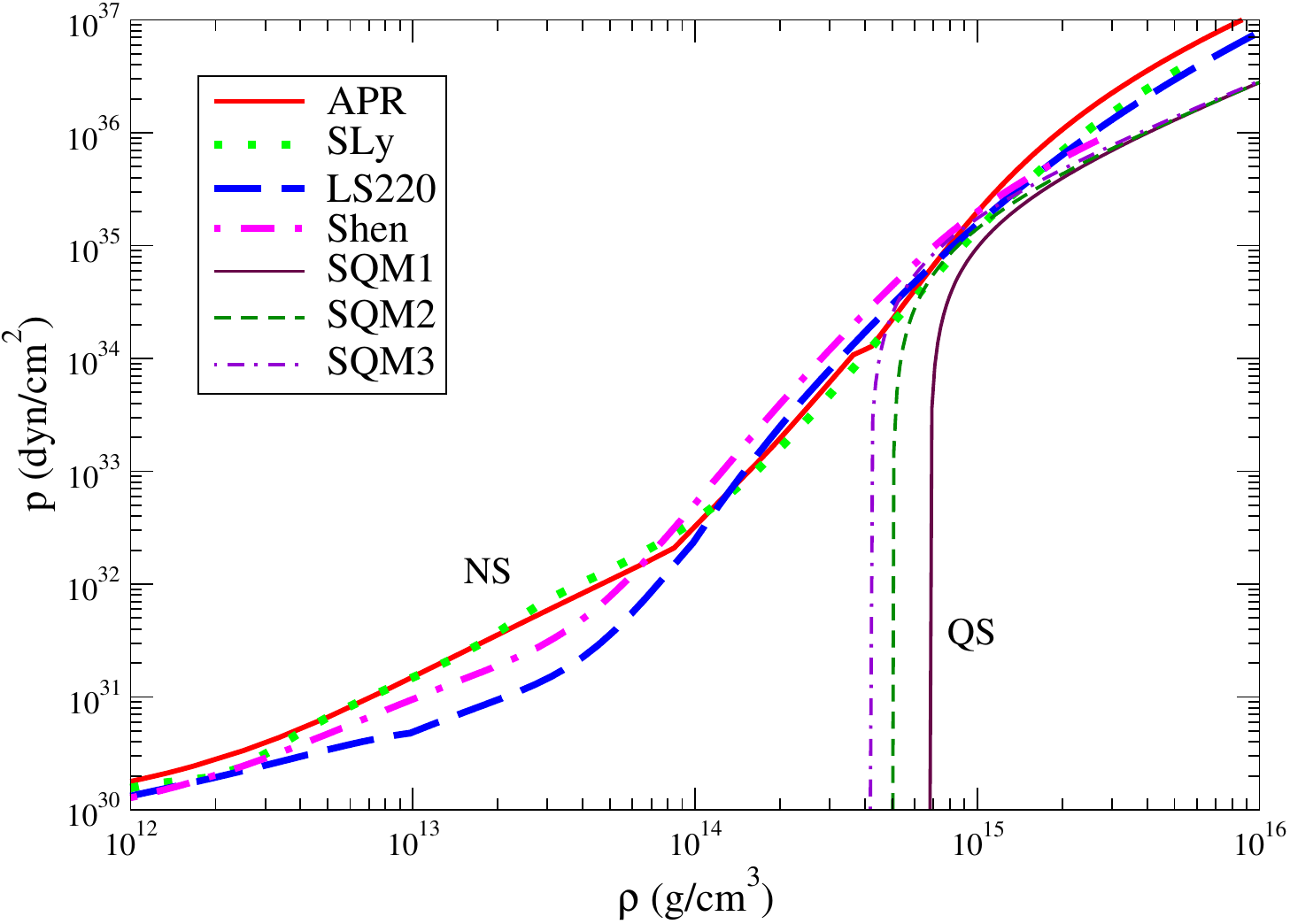}  
\caption{\label{fig:prho} 
(Color online) Various EoSs for NSs and QSs used in this paper.
}
\end{center}
\end{figure}

We consider four realistic EoSs for NSs: APR~\cite{APR}, SLy~\cite{SLy}, Lattimer and Swesty with nuclear incompressibility of 220MeV (LS220)~\cite{LS} and Shen~\cite{Shen1,Shen2}. For the latter two EoSs, we impose the neutrino-less and beta-equilibrium condition with the NS temperature of 0.1MeV. We also consider three EoSs for QSs: SQM1, SQM2 and SQM3~\cite{SQM}. The EoSs mentioned above are shown in Fig.~\ref{fig:prho}. For comparison, we also consider polytropic EoSs given by
\be
p = K \rho^{1+1/n}\,, 
\ee
where $K$ and $n$ are an amplitude constant and a polytropic index, respectively.

\begin{figure*}[htb]
\begin{center}
\includegraphics[width=8.5cm,clip=true]{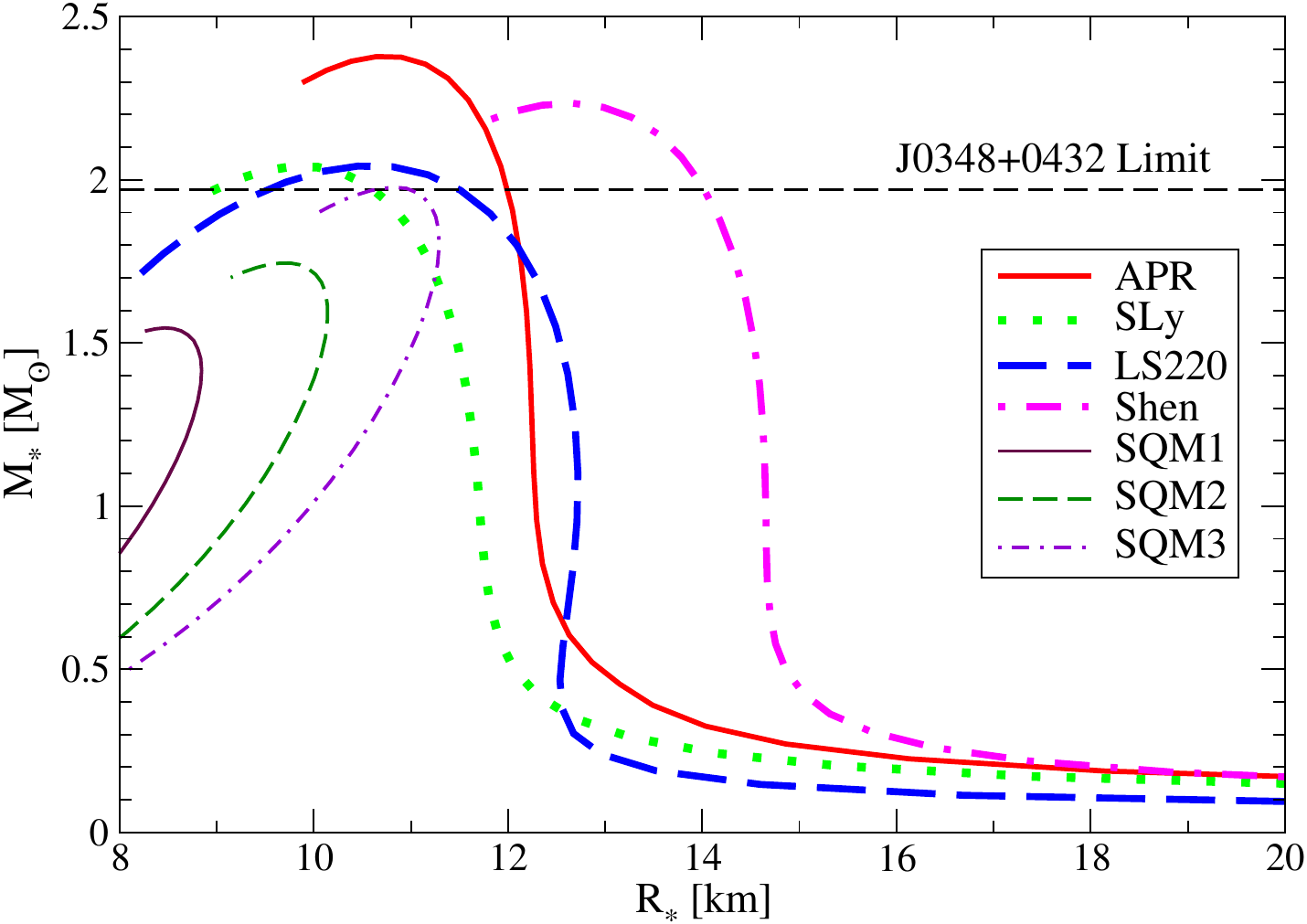}  
\includegraphics[width=8.5cm,clip=true]{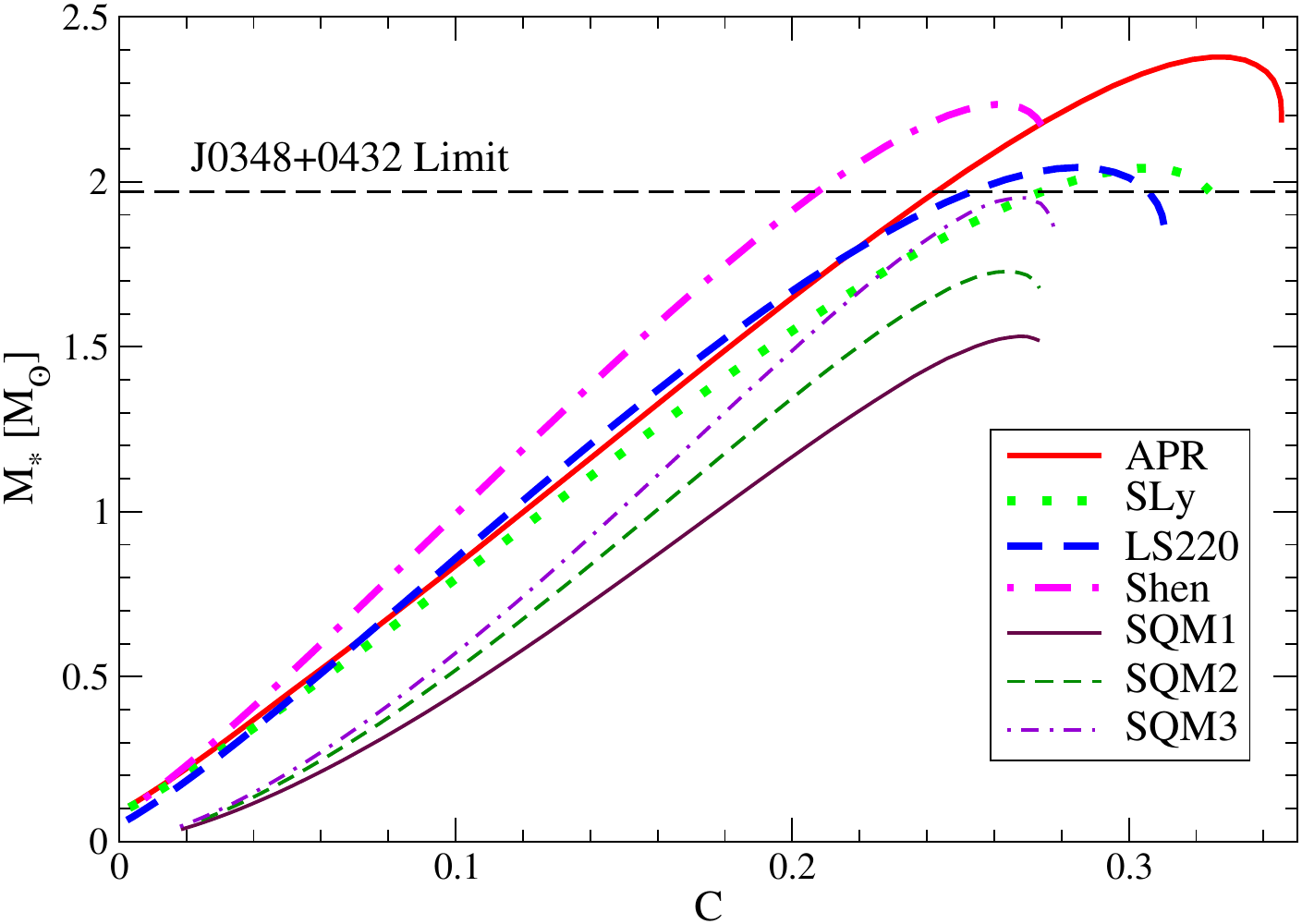}  
\caption{\label{fig:MR} 
(Color online) A mass-radius (left) and mass-compactness (right) relations for NSs and QSs with various EoSs. For a reference, we show a lower bound mass of a pulsar PSR J0348+0432~\cite{2.01NS}.
}
\end{center}
\end{figure*}

The left panel of Fig.~\ref{fig:MR} shows the mass-radius relations for various EoSs. For a reference, we show a lower bound mass of a pulsar PSR J0348+0432~\cite{2.01NS}. The right panel of Fig.~\ref{fig:MR} shows the mass-compactness ($C \equiv M_* / R_*$) relation.

\subsection{Perturbations}
\label{sec:pert}

Let us now consider perturbations to the spherically symmetric NS solution constructed in Sec.~\ref{sec:bg}. Thanks to the background symmetry, the angular sector of perturbations can be decomposed using (tensor) spherical harmonics. We restrict ourselves to stationary perturbations.

\subsubsection{Electric-Type}
\label{sec:E}

Let us first consider even-parity perturbations. The metric perturbation in the Regge-Wheeler gauge can be expressed as~\cite{thorne-campolattaro,ipser-price,lindblom}
\allowdisplaybreaks
\ba
ds^2 &=& ds_0^2 - \left[ e^{\nu (r)} H_{0,\ell} (r) Y_{\ell m} (\theta, \varphi) dt^2 \right. \nn \\
& & \left. + 2 H_{1,\ell} (r) Y_{\ell m} (\theta, \varphi) dt dr \right.  \nn \\
& & \left. + e^{\lambda (r)} H_{2,\ell} (r) Y_{\ell m} (\theta, \varphi) dr^2 \right. \nn \\
& & \left. +r^2 K_{\ell} (r) Y_{\ell m} (\theta, \varphi) (d \theta^2 + \sin^2 \theta d\varphi^2) \right]\,, \nn \\
\ea
while the matter perturbation is given by~\cite{thorne-campolattaro}
\allowdisplaybreaks
\ba
\delta T^0_0 (r, \theta, \varphi) &=& - \delta \rho_\ell (r) Y_{\ell m} (\theta, \varphi) \nn \\
& & = - \frac{d\rho}{dp} \delta p_\ell (r) Y_{\ell m} (\theta, \varphi)\,, \\
\delta T^i_i (r, \theta, \varphi) &=& \delta p_\ell (r) Y_{\ell m} (\theta, \varphi)\,.
\ea
By plugging these perturbed quantities into the field equations and the stress-energy conservation equations, one finds 
\ba
H_{0,\ell} &=& H_{2,\ell} \equiv H_\ell\,, \\
H_{1,\ell} &=& 0\,, \\
\label{eq:log-enthalpy}
\frac{\delta p_\ell}{\rho + p} &=& -\frac{1}{2}H_\ell\,, 
\ea
which means that we are only left with two perturbation parameters, $H_\ell$ and $K_\ell$. One can further use the perturbed field equations to eliminate $K_\ell$, which then leaves us with the master equation as~\cite{lindblom}
\ba
\label{eq-H0}
& & \frac{d^2 H_\ell}{dr^2} + \left\{ \frac{2}{r} + e^{\lambda} \left[ \frac{2M}{r^2} + 4 \pi r (p-\rho) \right] \right\} \frac{dH_\ell}{dr} \nn \\
& & + \left\{ e^\lambda \left[ - \frac{\ell (\ell +1)}{r^2} + 4 \pi (\rho + p) \frac{d\rho}{dp} + 4 \pi (5 \rho + 9 p) \right] \right. \nn \\
& & \left. - \left( \frac{d\nu}{dr} \right)^2 \right\} H_\ell   = 0\,.
\ea

\begin{figure}[htb]
\begin{center}
\includegraphics[width=8.5cm,clip=true]{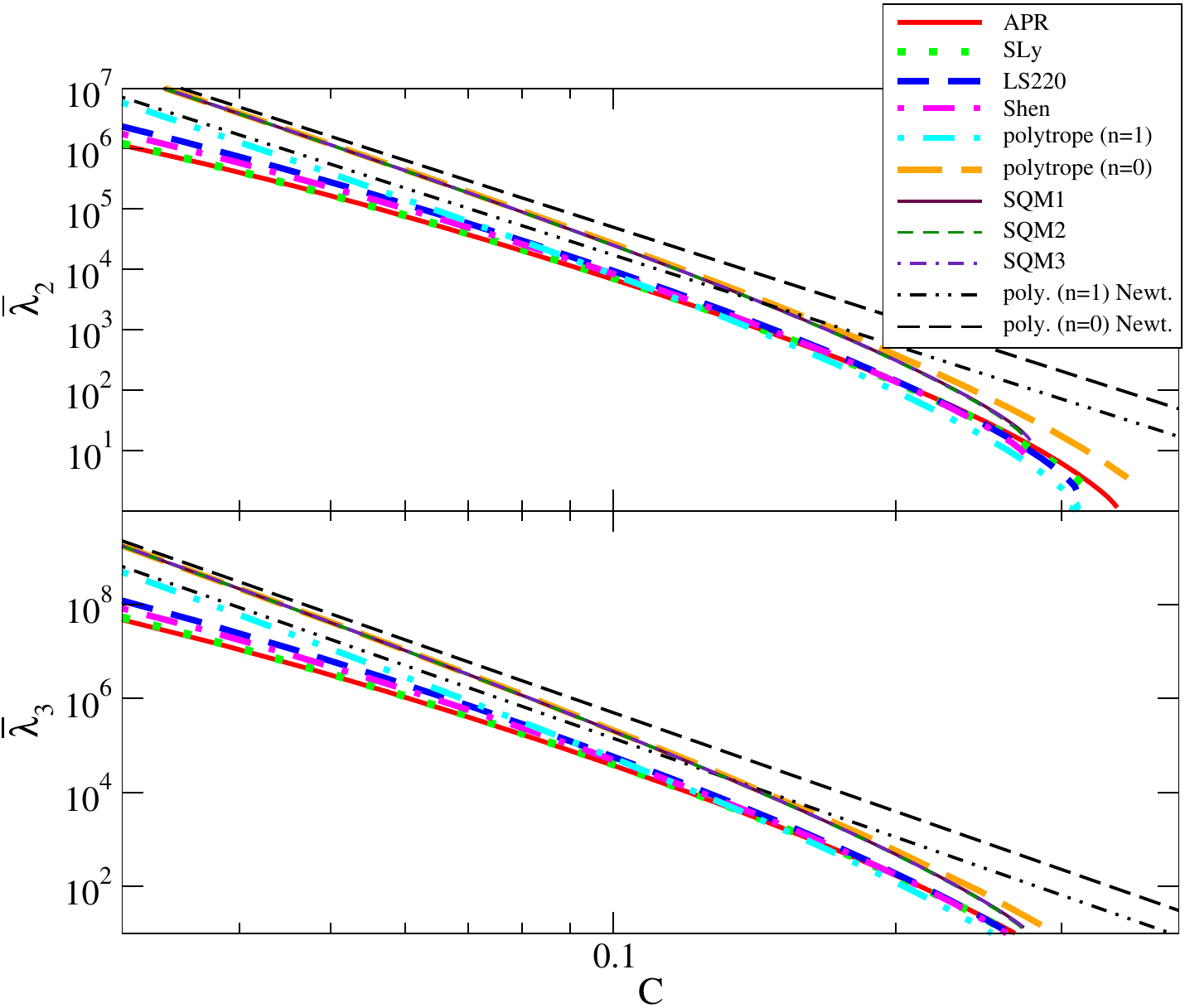}  
\caption{\label{fig:lambdabar-C} 
(Color online) $\bar{\lambda}_2$--$C$ (top) and $\bar{\lambda}_3$--$C$ (bottom) relations for NSs and QSs with various EoSs. We also show the Newtonian relations for the $n=1$ and $n=0$ polytropes. Observe that the QS relations are similar to those for the $n=0$ polytrope for $C<0.2$, which indicates that tidal deformabilities are most sensitive to the star's outer layer.
}
\end{center}
\end{figure}

One first solves the above differential equation with an initial condition $H_\ell (r) \propto r^\ell$, where we have imposed regularity at the NS center and the constant of proportionality is irrelevant in further calculations of the tidal deformabilities. On the other hand, the exterior solution is obtained as 
\be
H_\ell = a_{\ell}^{P} \hat{P}_{\ell 2}(x) + a_{\ell}^{Q} \hat{Q}_{\ell 2}(x)\,,
\ee
where $x \equiv r/M_* -1$ and $\hat{P}_{\ell 2}(x)$ and $\hat{Q}_{\ell 2}(x)$ are the normalized associated Legendre functions of the first and second kinds respectively, with $\hat{P}_{\ell 2}(x) \sim x^{\ell}$ and $\hat{Q}_{\ell 2}(x) \sim x^{-(\ell + 1)}$ for $x \rightarrow \infty$. $a_{\ell}^{P}$ and $a_{\ell}^{Q}$ are constants whose ratio is to be determined by matching the logarithmic derivative of the interior and exterior solutions, 
\be
\label{eq:y-lambda}
y_\ell^\lambda (r) \equiv \frac{r}{H_\ell(r)} \frac{dH_\ell(r)}{dr}\,,
\ee
at the NS surface\footnote{One must be careful when performing the matching for a constant density star or a QS due to the density discontinuity at the surface, as explained in~\cite{damour-nagar,hinderer-lackey-lang-read}. For such stars, $y_\ell^\lambda (R)$ acquires an extra term that depends on the mass, radius and the interior density at the surface.}, which yields
\be
\label{al}
a_{\ell}^{\lambda} \equiv \frac{a_\ell^Q}{a_\ell^P} = - \frac{\hat{P}'_{\ell 2}(x_c) - C y_\ell^\lambda (R_*) \hat{P}_{\ell 2}(x_c)}{\hat{Q}'_{\ell 2}(x_c) - C y_\ell^\lambda (R_*) \hat{Q}_{\ell 2}(x_c)}\,,
\ee
where $'$ denotes the derivative with respect to $x$, $x_c \equiv C^{-1} -1$ and we recall that $C$ is the NS compactness. One can show that $a_\ell^\lambda$ is related to $\lambdabarnew_\ell$ by~\cite{damour-nagar}
\be
\label{lambdabar-l}
\lambdabarnew_\ell = \frac{a_\ell^\lambda}{(2 \ell - 1)!!}\,.
\ee
In Fig.~\ref{fig:lambdabar-C}, we plot $\lambdabarnew_2$ and $\lambdabarnew_3$ against the compactness $C$ for various EoSs. Similar relation holds for $\lambdabarnew_4$. Observe that the NS relations vary from one EoS to another, while the QS relations are almost identical among the EoS considered here. Also, notice that the QS relations are similar to that of the $n=0$ polytrope for $C<0.2$. Since the outer layer of a QS behaves as a constant density ($n=0$ polytropic) star, this figure shows an evidence that the tidal deformabilities are most sensitive to the star's outer layer, as discussed in~\cite{I-Love-Q-Science,I-Love-Q-PRD}.

Let us now derive $\lambdabarnew_\ell$ in the Newtonian limit for the $n=1$ and $n=0$ polytropes. Eqs.~\eqref{eq-H0} and~\eqref{lambdabar-l} in this limit becomes 
\be
\frac{d^2 H_\ell^\N}{dr^2} + \frac{2}{r} \frac{d H_\ell^\N}{dr} - \left( \frac{\ell (\ell + 1)}{r^2}  - 4 \pi \rho \frac{d\rho}{dp} \right) H_\ell^\N = 0\,,
\ee
and
\be
\label{lambda-l-N}
\lambdabarnew_\ell^\N = \frac{1}{(2 \ell -1)!!} \frac{\ell - y_\ell^\lambda}{\ell + 1 - y_\ell^\lambda} \frac{1}{C^{2 \ell + 1}}\,,
\ee
respectively.
For the $n=1$ polytropes, the solution that is regular at the NS center is given by Bessel functions as\footnote{There is a typo in~\cite{I-Love-Q-PRD,hinderer-love}.} $H_\ell^\N{}^{, (n=1)} \propto \sqrt{R_*/r} J_{\ell + 1/2} (\pi r/R_*)$. From this solution, one can calculate $y_\ell^\lambda$ in the Newtonian limit and by plugging it into Eq.~\eqref{lambda-l-N}, one obtains $\lambdabarnew_\ell$ for $\ell = 2,3,4$ in the Newtonian limit as
\ba
\label{eq:lambdabar-Newtonian-n1-2}
\lambdabarnew_2^\N{}^{, (n=1)} &=& \frac{15-\pi^2}{3 \pi^2} \frac{1}{C^5}\,, \\
\label{eq:lambdabar-Newtonian-n1-3}
\lambdabarnew_3^\N{}^{, (n=1)} &=& \frac{21-2 \pi^2}{9 \pi^2} \frac{1}{C^7}\,, \\
\label{eq:lambdabar-Newtonian-n1-4}
\lambdabarnew_4^\N{}^{, (n=1)} &=& -\frac{945 - 105 \pi^2 + \pi^4}{105 \pi^2 (\pi^2 -15)} \frac{1}{C^9}\,. 
\ea
For the $n=0$ polytropes, $\lambdabarnew_\ell$ in the Newtonian limit is given by~\cite{damour-nagar}
\be
\label{eq:lambdabar-Newtonian-n0}
\lambdabarnew_\ell^\N{}^{, (n=0)} = \frac{3}{2 (\ell -1) (2 \ell -1)!!} \frac{1}{C^{2 \ell +1}}\,.
\ee
These Newtonian relations are shown as black dotted-dashed ($n=1$) and dashed ($n=0$) lines in Fig.~\ref{fig:lambdabar-C}.
Notice that the relativistic relations for such polytropes approach the Newtonian lines as one decreases $C$.

\subsubsection{Magnetic-Type}

Next, we consider magnetic-type tidal perturbations. The odd-parity metric stationary perturbation is given by~\cite{thorne-campolattaro}
\ba
ds^2 &=& ds_0^2 - 2 h_{0,\ell} (r) \frac{\partial_\varphi Y_{\ell m} (\theta, \varphi)}{\sin \theta}  dt d\theta \nn \\
& & +2 h_{0,\ell} (r) \sin \theta \partial_\theta Y_{\ell m} (\theta, \varphi) dt d\varphi\,. 
\ea
The matter stress-energy tensor is perturbed only through the metric. The master equation is obtained as~\cite{cunningham,andrade}
\ba
\label{eq:d2hdr2}
&  & \frac{d^2 h_\ell}{dr^2} + \frac{e^{\lambda}}{r^2} \left[ 2 M + 4 \pi (p - \rho) r^3 \right] \frac{d h_\ell}{dr} \nn \\
& & - e^{\lambda} \left[ \frac{\ell (\ell + 1)}{r^2} - \frac{6 M}{r^3} + 4 \pi (\rho - p) \right] h_\ell = 0\,,
\ea
where the master variable $h_\ell (r)$ is defined by 
\be
h_\ell (r) \equiv r^3 \partial_r \left( \frac{h_{0,\ell}}{r^2} \right)\,.
\ee

\begin{figure}[thb]
\begin{center}
\includegraphics[width=8.5cm,clip=true]{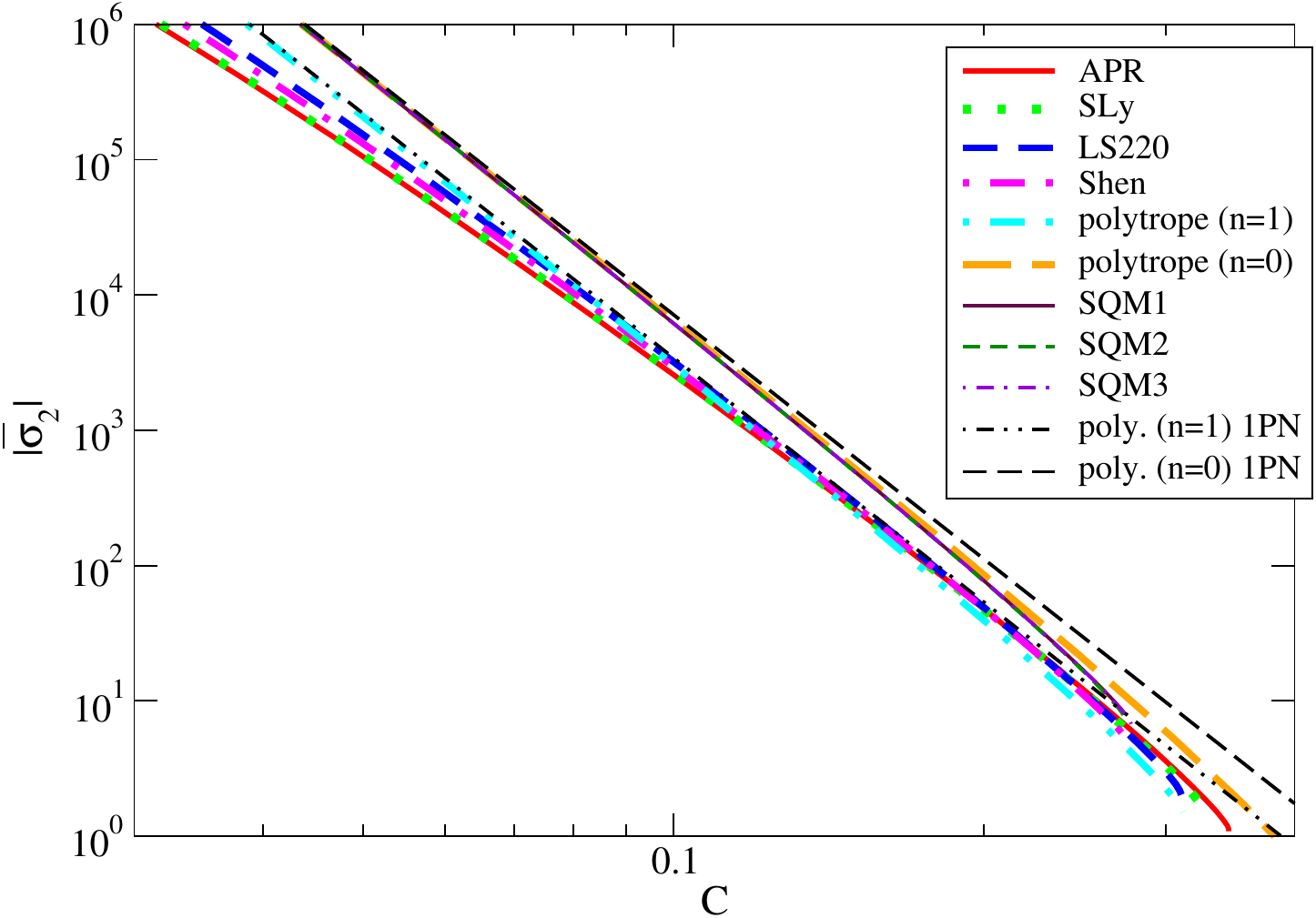}  
\caption{\label{fig:sigmabar-C} 
(Color online) $\bar{\sigma}_2$--$C$ relation for NSs and QSs with various EoSs. We also show the relations for the $n=1$ and $n=0$ polytropes at 1PN order.
}
\end{center}
\end{figure}

One can follow a similar procedure to Sec.~\ref{sec:E} to calculate $\sigmabar_\ell$~\cite{damour-nagar}. One first needs to solve Eq.~\eqref{eq:d2hdr2} in the NS interior with the initial condition given by $h_\ell \propto r^{\ell + 1}$, where we have imposed regularity at the NS center. Again, the constant of proportionality is irrelevant. For the exterior solution, the asymptotic behavior of the two independent solutions at spatial infinity is given by $\hat{h}_\ell^P (\hat{r}) \sim \hat{r}^{\ell +1}$ and $\hat{h}_\ell^Q (\hat{r}) \sim \hat{r}^{-\ell}$, where $\hat{r} \equiv r/M_*$. In particular, for $\ell = 2$, the exterior solutions are given by $\hat{h}_2^P (\hat{r}) = \hat{r}^3$ and $\hat{h}_2^Q (\hat{r}) = - \hat{r}^3 \partial_{\hat{r}} [ F(1,4;6;2/\hat{r})/\hat{r}^4]/4$, where $F(a,b;c;z)$ is a hypergeometric function.  Then, the general exterior solution is given by
\be
h_\ell (r) = b_\ell^P \hat{h}_2^P (\hat{r}) + b_\ell^Q \hat{h}_2^Q (\hat{r})\,,
\ee
where, again, the ratio of the coefficients $b_\ell^P$ and $b_\ell^Q$ are determined by matching $y_\ell^\sigma (\hat{r}) \equiv (r/h_\ell) (d h_\ell/dr)$ at the NS surface, which yields
\be
b_\ell \equiv \frac{b_\ell^Q}{b_\ell^P} = - \frac{\hat{h}_\ell^P{}'(C^{-1}) - C y_\ell^\sigma (C^{-1}) \hat{h}_\ell^P (C^{-1})}{\hat{h}_\ell^Q{}'(C^{-1}) - C y_\ell^\sigma (C^{-1}) \hat{h}_\ell^Q (C^{-1})}\,.
\ee
One can show that $\sigmabar_\ell$ is obtained from $b_\ell$ via~\cite{damour-nagar}
\be
\label{sigmabar-l}
\sigmabar_\ell = \frac{\ell -1}{4 (\ell +2) (2 \ell -1)!!} b_\ell\,.
\ee
In Fig.~\ref{fig:sigmabar-C}, we plot $\sigmabar_2$ against $C$ for various EoSs.

Next, let us derive $\sigmabar_2$ in the Newtonian limit~\cite{damour-nagar}. Eqs.~\eqref{eq:d2hdr2} and~\eqref{sigmabar-l}  in this limit are given by
\be
\label{eq:sigma-N}
\frac{d^2 h_\ell}{dr^2} = \frac{\ell (\ell + 1)}{r^2} h_\ell\,,
\ee
and
\be
\sigmabar_\ell^\N =  \frac{\ell -1}{4 (\ell +2) (2 \ell -1)!!} \frac{\ell + 1 - y_\ell^\sigma}{\ell + y_\ell^\sigma} \frac{1}{C^{2 \ell +1}}\,,
\ee
respectively.
Since Eq.~\eqref{eq:sigma-N} does not depend on $p$ nor $\rho$, the solution that is regular at the NS center has the form $h_\ell^\N \propto r^{\ell +1}$ both the NS interior and exterior, which means that $b_\ell^Q =0$, leading to $\sigmabar_\ell^\N = 0$. This is consistent with the fact that there is no Newtonian analogue for the magnetic-type tidal deformabilities. Going to 1PN order, one finds~\cite{damour-nagar}
\ba
y_\ell^\sigma (R_*) &=& (\ell +1) \left\{ 1 + \frac{\ell +2}{\ell +1} \frac{1}{R_*^{2 \ell +1}} \int_0^{R_*} dr r^{2\ell} \right. \nn \\
& & \left. \times \left[ 2(\ell -2) \frac{M}{r} + 4 \pi (\rho -p) r^2 \right] \right\}\,,
\ea
where the first and second terms correspond to the Newtonian and 1PN terms, respectively. In particular, $\sigmabar_2$ to 1PN order is given by 
\be
\sigmabar_2^\firstPN = - \frac{\pi}{15} \frac{1}{M_*^5} \int_0^{R_*} dr (\rho -p) r^6\,.  
\ee
For the $n=1$ and $n=0$ polytropes, $\sigmabar_2^\firstPN$ is given by~\cite{damour-nagar}
\ba
\label{eq:sigmabar-Newtonian-n1}
\sigmabar_2^{\firstPN, (n=1)} &=& - \frac{1}{60} \left( 1 - \frac{20}{\pi^2} + \frac{120}{\pi^4} \right) \frac{1}{C^6}\,, \\
\label{eq:sigmabar-Newtonian-n0}
\sigmabar_2^{\firstPN, (n=0)} &=& - \frac{1}{140} \frac{1}{C^6}\,, 
\ea
respectively.
These relations at 1PN order are also shown in Fig.~\ref{fig:sigmabar-C}.

\subsubsection{Shape}

Finally, we will calculate the shape tidal deformability $\etabar_\ell$. From the fact the the logarithmic enthalpy $\bar{h} \equiv \int^p_0 dp/(\rho + p)$ vanishes at the NS surface and using Eqs.~\eqref{TOV-zeroth} and~\eqref{eq:log-enthalpy}, one finds~\cite{damour-nagar}
\be
\label{deltaRdR1}
\frac{\delta R_\ell}{R_*} = - \frac{1}{2} \left[ \frac{1-2C}{C} H_\ell (R_*) + K_\ell (R_*) \right]\,. 
\ee
In the Regge-Wheeler gauge, $K_\ell (R_*)$ can be obtained from a linear combination of $H_\ell (R_*)$ and $H_\ell '(R_*)$ as~\cite{lindblom}
\be
\label{eq:K}
K_\ell (R_*) = \alpha_1 R_* H_\ell '(R_*) + \alpha_2 H_\ell (R_*)\,,
\ee
with
\ba
\label{eq:alpha1}
\alpha_1 &=& \frac{2 C}{(\ell -1) (\ell +2) }\,, \\
\label{eq:alpha2}
\alpha_2 &=& \frac{1}{(\ell -1) (\ell +2)} \left[ \ell (\ell +1) + \frac{4 C^2}{1-2 C} - 2 (1-2C) \right]\,. \nn \\
\ea
By substituting Eq.~\eqref{eq:K} into Eq.~\eqref{deltaRdR1} and using Eq.~\eqref{eq:y-lambda}, one obtains
\be
\label{deltaRdR2}
\frac{\delta R_\ell}{R_*} = - \frac{H_\ell (R_*)}{2} \left( \frac{1-2C}{C} + y_\ell^\lambda (R_*) \alpha_1 + \alpha_2 \right)\,. 
\ee

Following~\cite{damour-nagar}, we define the $\ell$-th component of the disturbing potential $U_\ell (R_*)$ to be the leading asymptotically growing part in $-H_\ell/2$ at $r=R_*$, i.e.
\be
\label{disturbing}
U_\ell (R_*) = - \frac{1}{2} \frac{a_\ell^P}{C^\ell}\,. 
\ee
Rewriting $a_\ell^P$ and $a_\ell^Q$ in terms of $U_\ell$ and $a_\ell \equiv a_\ell^Q/a_\ell^P$ using the above equation, one finds
\be
- \frac{1}{2} H(R_*) = C^\ell \hat{P}_{\ell 2} (x) \left[ 1 + a_\ell \frac{\hat{Q}_{\ell 2} (x)}{\hat{P}_{\ell 2} (x)} \right] U_\ell (R_*)\,.
\ee
From Eqs.~\eqref{eq:shape},~\eqref{deltaRdR2} and~\eqref{disturbing}, one finally obtains
\ba
\etabar_\ell &=& \frac{2}{(2 \ell -1)!!} \frac{\hat{P}_{\ell 2} (x_c)}{C^\ell} \left( \frac{1-2C}{C} + y_\ell^\lambda \alpha_1 + \alpha_2 \right) \nn \\
& & \times \left( 1-\frac{\partial_x \ln \hat{P}_{\ell 2} (x_c) - C y_\ell^\lambda (R_*)}{\partial_x \ln \hat{Q}_{\ell 2} (x_c) - C y_\ell^\lambda (R_*)} \right) \nn \\
&=& \left\{ \left[ 2 \lambdabarnew_\ell \hat{Q}_{\ell 2} (x_c)  + \frac{2}{(2 \ell -1)!!} \hat{P}_{\ell 2} (x_c) \right]  [1+(\alpha_2 -2) C] \right. \nn \\
& & \left. + \left[ 2 \lambdabarnew_\ell \hat{Q}'_{\ell,2}(x_c) + \frac{2}{(2 \ell -1)!!} \hat{P}'_{\ell,2}(x_c) \right] \alpha_1 \right\} \frac{1}{C^{\ell+1}}\,, \nn \\
\ea
where we used Eqs.~\eqref{al} and~\eqref{lambdabar-l} to rewrite $y_\ell^\lambda (R_*)$ in terms of $\lambdabarnew_\ell$.
In the Newtonian limit, one finds
\be
\label{eq:etabar-Newtonian}
\etabar_\ell^\N = 2 \lambdabarnew_\ell^\N + \frac{2}{(2 \ell -1)!!} \frac{1}{C^{2\ell +1}}\,.
\ee

\begin{figure}[htb]
\begin{center}
\includegraphics[width=8.5cm,clip=true]{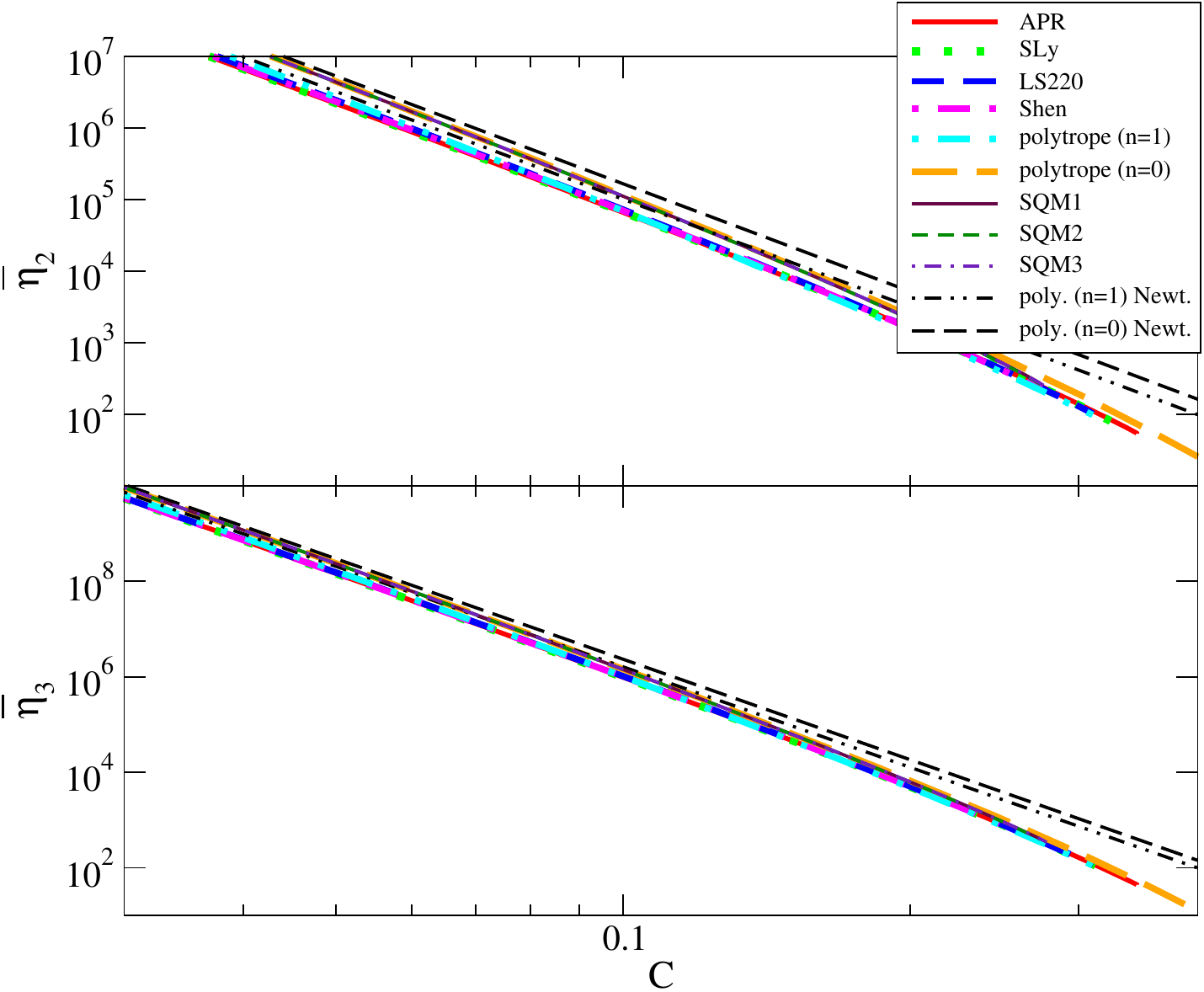}  
\caption{\label{fig:lambdabar-shape-C} 
(Color online) $\bar{\eta}_2$--$C$ (top) and $\bar{\eta}_3$--$C$ (bottom) relations for NSs and QSs with various EoSs. We also show the Newtonian relations for the $n=1$ and $n=0$ polytropes.
}
\end{center}
\end{figure}

Figure~\ref{fig:lambdabar-shape-C} shows $\bar{\eta}_2$ and $\bar{\eta}_3$ against $C$ for various EoSs, together with the Newtonian lines for the $n=1$ and the $n=0$ polytropes. Similar relation holds for $\bar{\eta}_4$. Notice that NS and QS curves almost lie on top of each other. This is different from the $\lambdabarnew_\ell$--$C$ and $\sigmabar_\ell$--$C$ relations shown in Figs.~\ref{fig:lambdabar-C} and~\ref{fig:sigmabar-C}. Figure~\ref{fig:lambdabar-shape-C} already indicates that the universal relations exist among $\etabar_\ell$, which we will discuss next.

\section{Universal Relations}
\label{sec:universal}

In this section, we first plot $\lambdabarnew_\ell$, $\sigmabar_\ell$ and $\etabar_\ell$ calculated in the previous section against each other to show the universal relations. Next, we carry out an analytic analysis to show that the universal relation roughly holds for the $n=1$ and $n=0$ polytropes in the Newtonian limit.


\begin{figure*}[htb]
\begin{center}
\includegraphics[width=8.5cm,clip=true]{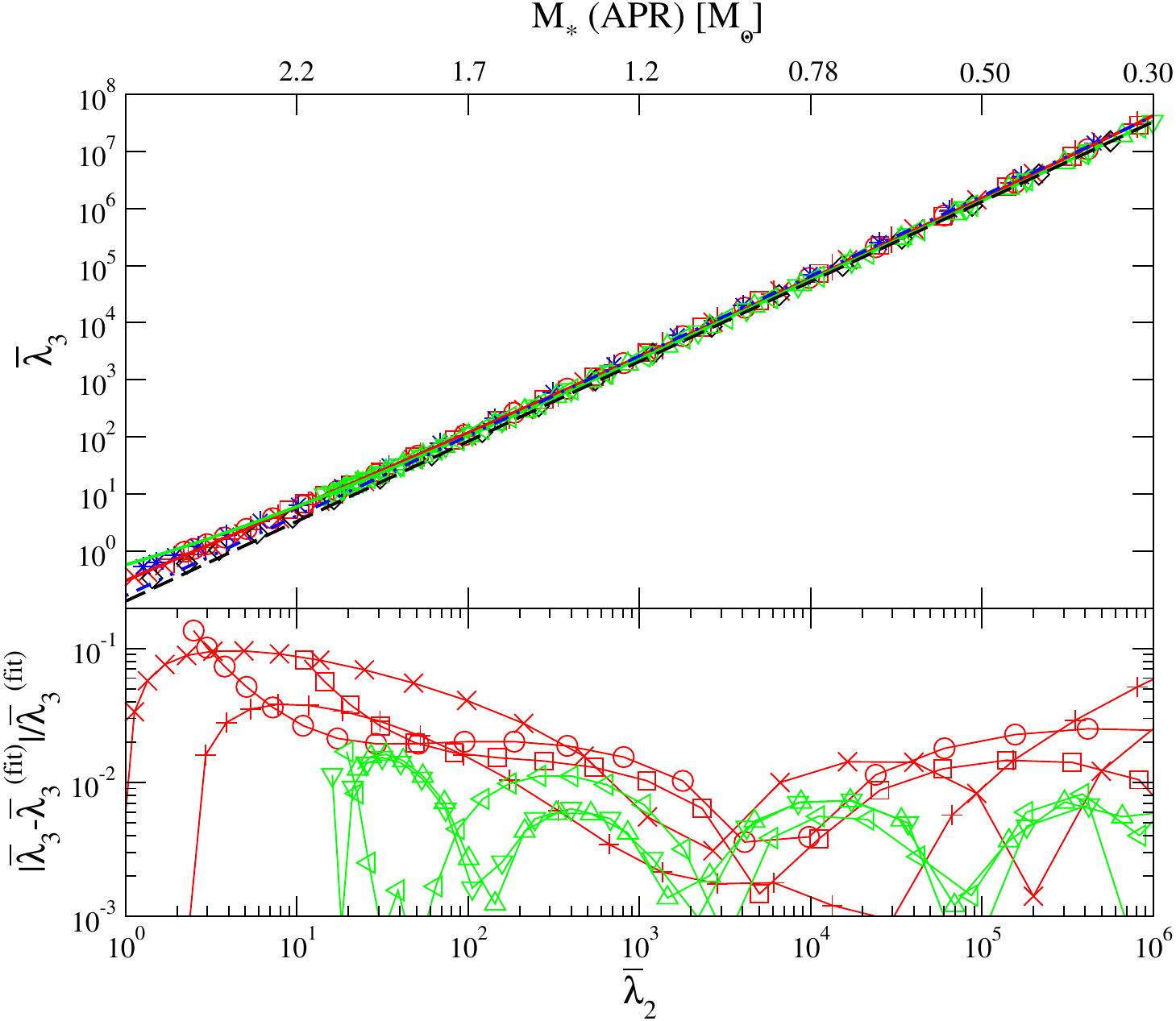}  
\includegraphics[width=8.5cm,clip=true]{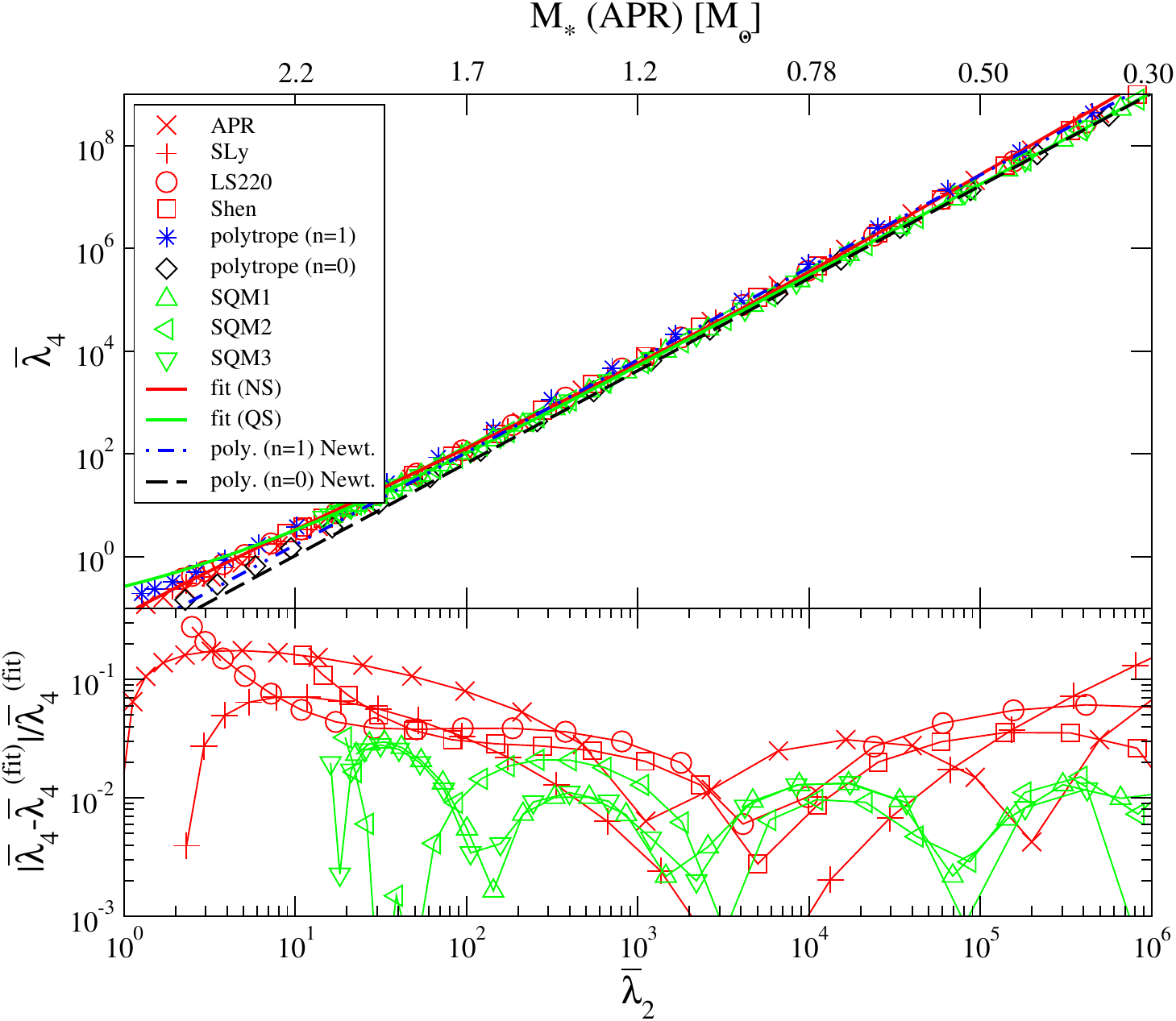}  
\caption{\label{fig:lambdabar34-lambdabar2} 
(Color online) (Top) Universal $\bar{\lambda}_3$--$\bar{\lambda}_2$ (left) and $\bar{\lambda}_4$--$\bar{\lambda}_2$ (right) relations for NSs (red) and QSs (green) with various realistic EoSs. We also show the relations with the $n=1$ (blue) and $n=0$ (black) polytropes, together with their Newtonian limit. The single parameter along the curves is the mass or compactness. We construct fitting formulas for NSs and QSs with the realistic EoSs given by Eq.~\eqref{fit}, which are shown with solid curves. As a reference, we show the NS mass for the APR EoS on the top axis.
(Bottom) Fractional difference of each curve to the fitting formulas. 
}
\end{center}
\end{figure*}

\begin{figure}[htb]
\begin{center}
\includegraphics[width=8.5cm,clip=true]{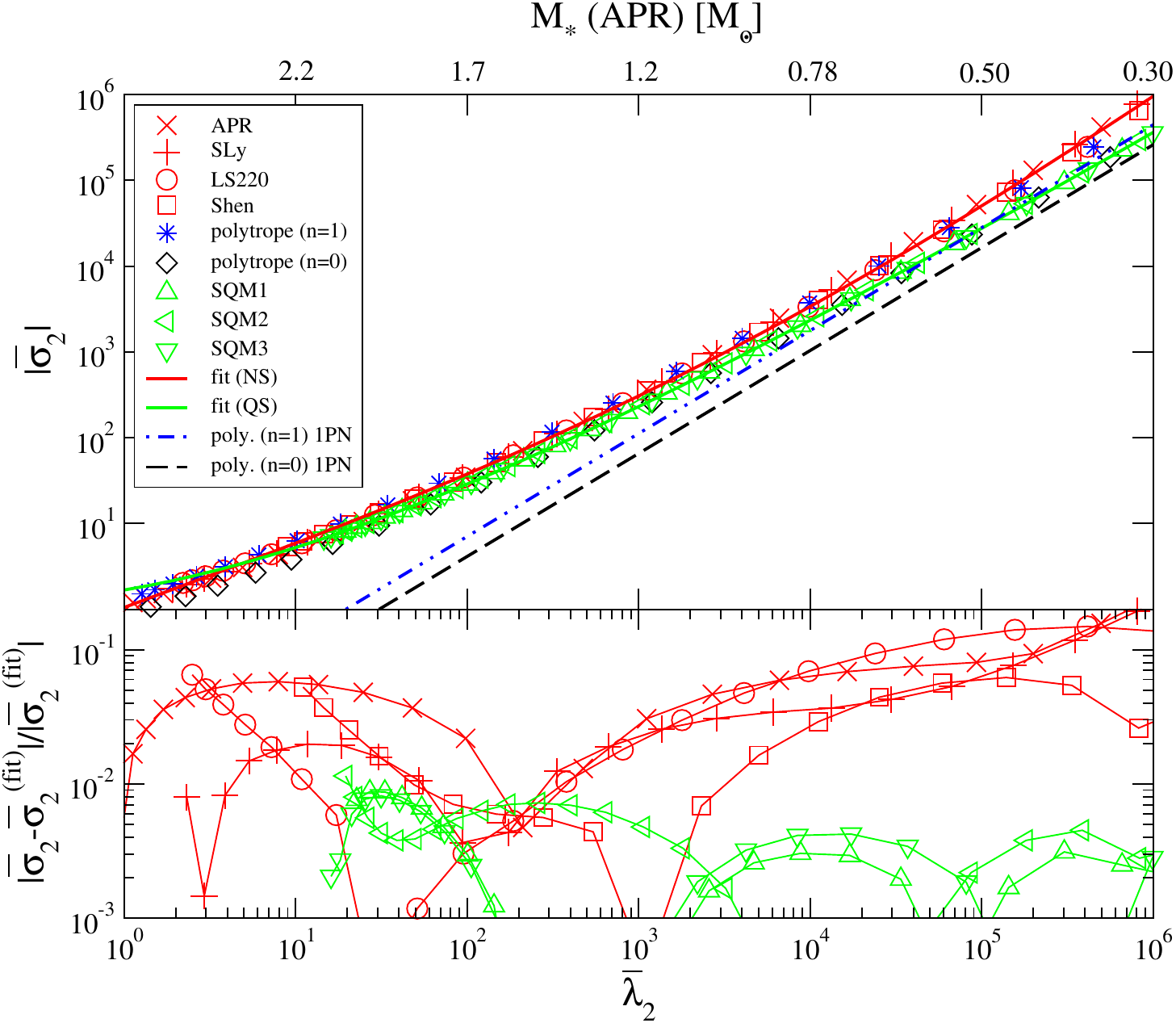}  
\caption{\label{fig:lambdabar2mag-lambdabar2} 
(Color online) (Top) Universal $|\bar{\sigma}_2|$--$\bar{\lambda}_2$ relation for NSs and QSs with various EoSs. The meaning of each curve is same as in Fig.~\ref{fig:lambdabar34-lambdabar2}. Observe that the NS relations are similar to that of the $n=1$ polytrope, while the QS relations are similar to that of the $n=0$ polytrope.
(Bottom) Fractional difference of each curve to the fitting formula. 
}
\end{center}
\end{figure}

The top panels of Figs.~\ref{fig:lambdabar34-lambdabar2} and~\ref{fig:lambdabar2mag-lambdabar2} show the universal $\bar{\lambda}_3$--$\bar{\lambda}_2$, $\bar{\lambda}_4$--$\bar{\lambda}_2$ and $\bar{\sigma}_2$--$\bar{\lambda}_2$ relations for NSs (red) and QSs (green) with various realistic EoSs. We also show the relations for the $n=1$ (blue) and $n=0$ (black) polytropes. We show the NS mass for the APR EoS on the top axis as a reference. The single parameter that characterizes the curves is the mass or compactness. One sees that the relations do not depend strongly on the EoS. We construct a fitting function (including only the realistic EoSs) of the form
\be
\ln y_i = a_i + b_i \ln x_i + c_i (\ln x_i)^2 + d_i (\ln x_i)^3+ e_i (\ln x_i)^4\,,
\label{fit}
\ee
where the fitted coefficients are shown in Table~\ref{table:coeff}. To include the accuracy of the fit, we construct the NS and QS fit separately. The fitted curves are shown as solid ones in each top panel of the figures. In the bottom panels, we show a fractional difference between the numerical and the fitted values. One sees that the universality holds to $\mathcal{O}(1)$--$\mathcal{O}(10)$\%. Among the EoSs considered here, the QS relations are tighter than those for NSs. Observe that as one increases the compactness, both the NS and QS branches approach the black hole (BH) limit~\cite{damour-nagar,binnington-poisson,fang-lovelace,kol-smolkin,chakrabarti1}, $\lambdabarnew_{\ell}^\BH = 0 = \sigmabar_{\ell}^{\BH}$.
Observe also that the NS relations are similar to that of the $n=1$ polytrope, while those for QSs are similar to that of the $n=0$ polytrope. The latter is expected since the outer layer of a QS behaves as a constant density star.

\begin{figure*}[htb]
\begin{center}
\includegraphics[width=8.5cm,clip=true]{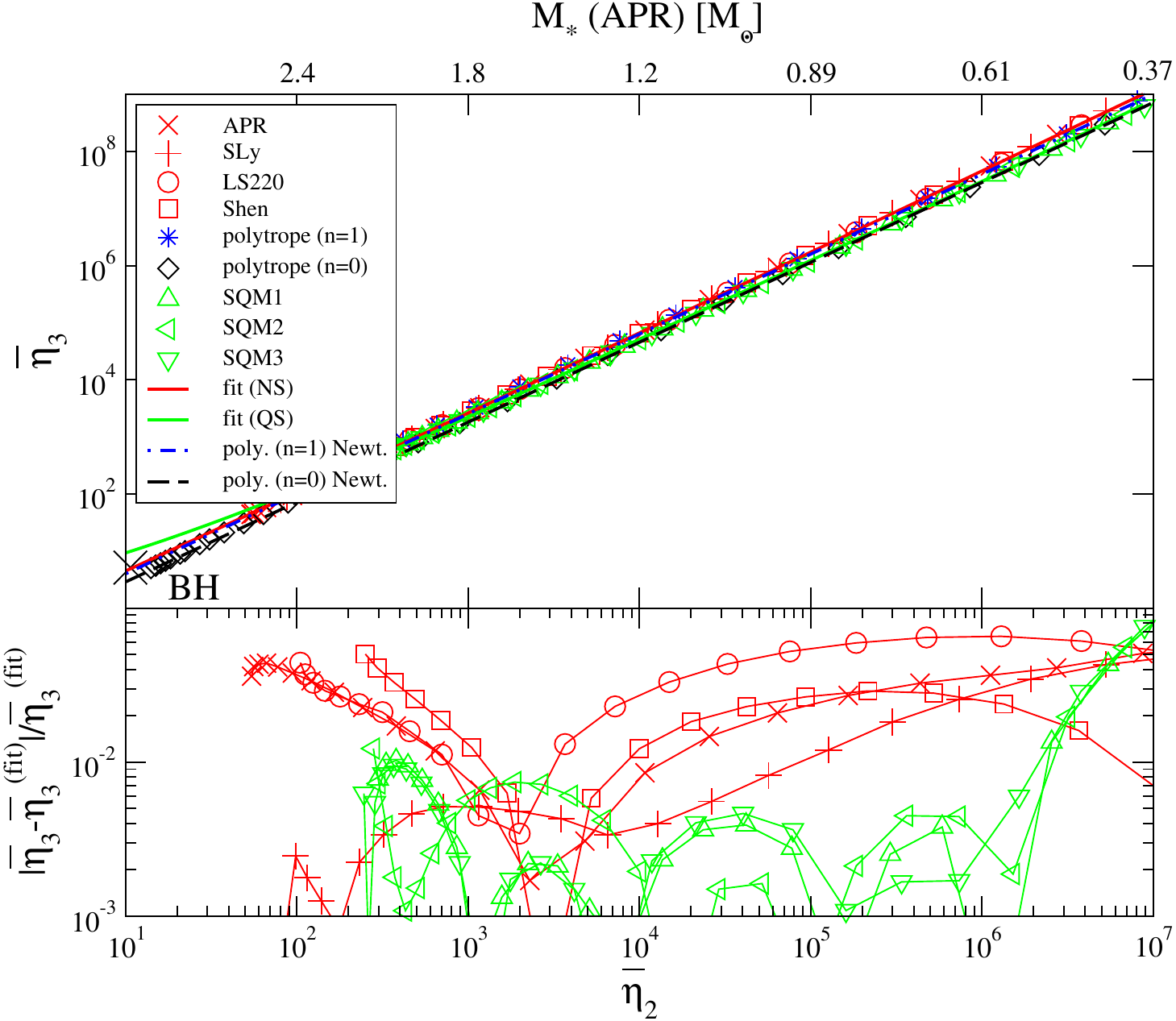}  
\includegraphics[width=8.5cm,clip=true]{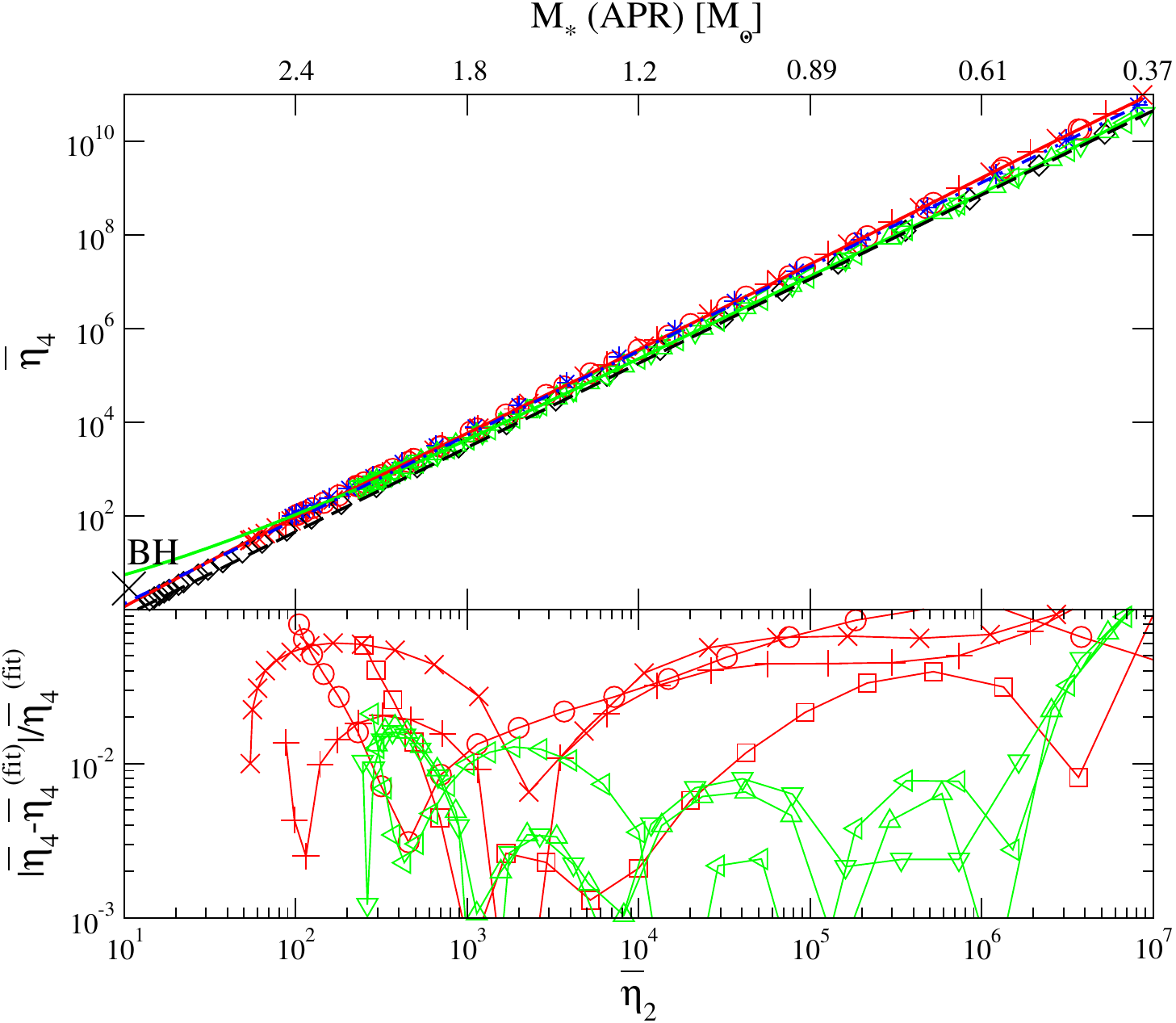}  
\caption{\label{fig:lambdabar34-lambdabar2-shape} 
(Color online) (Top) Universal $\bar{\eta}_3$--$\bar{\eta}_2$ (left) and $\bar{\eta}_4$--$\bar{\eta}_2$ (right) relations for NSs and QSs with various EoSs. The meaning of each curve is same as in Fig.~\ref{fig:lambdabar34-lambdabar2}. The big black cross shows the non-spinning BH values given by Eq.~\eqref{etabar-BH}. Observe that the Newtonian curves give a good approximation throughout.
(Bottom) Fractional difference of each curve to the fitting formulas. 
}
\end{center}
\end{figure*}

The top panels of Fig.~\ref{fig:lambdabar34-lambdabar2-shape} shows the universal $\bar{\eta}_3$--$\bar{\eta}_2$ and $\bar{\eta}_4$--$\bar{\eta}_2$ relations. We again construct fitting functions given in Eq.~\eqref{fit} with the coefficients shown in Table~\ref{table:coeff}. The fractional difference between the numerical and fitted values are shown in the bottom panel of Fig.~\ref{fig:lambdabar34-lambdabar2-shape}. The BH shape Love numbers are calculated in~\cite{lecian}, which then leads to the BH shape tidal deformabilities as 
\be
\etabar_\ell^\BH = \frac{2^{2\ell +1} (\ell +1)! (\ell !)^2}{(2 \ell -1)!! (2 \ell)! (\ell -1)}\,,
\label{etabar-BH}
\ee
where we used $C^\BH = 1/2$. In particular, $\etabar_2^\BH \approx 10.7$, $\etabar_3^\BH \approx 5.12$ and $\etabar_4^\BH \approx 2.79$, which are shown as big black crosses in Fig.~\ref{fig:lambdabar34-lambdabar2-shape}. Observe that the universal relations approach the BH limit as one increases the compactness.

Next, we follow~\cite{I-Love-Q-PRD} and we derive analytic relations for the $n=1$ and the $n=0$ polytropes in the Newtonian limit. From Eqs.~\eqref{eq:lambdabar-Newtonian-n1-2}--\eqref{eq:lambdabar-Newtonian-n0},~\eqref{eq:sigmabar-Newtonian-n1},~\eqref{eq:sigmabar-Newtonian-n0} and~\eqref{eq:etabar-Newtonian}, one derives
\ba
\allowdisplaybreaks
\lambdabarnew_3^\N &=& C_{\lambdabarnew_3 \lambdabarnew_2} \left( \lambdabarnew_2^N \right)^{7/5}\,, \\
\lambdabarnew_4^\N &=& C_{\lambdabarnew_4 \lambdabarnew_2} \left( \lambdabarnew_2^N \right)^{9/5}\,, \\
\sigmabar_2^\firstPN &=& C_{\sigmabar_2 \lambdabarnew_2} \left( \lambdabarnew_2^N \right)^{6/5}\,, \\
\etabar_3^\N &=& C_{\etabar_3 \etabar_2} \left( \etabar_2^N \right)^{7/5}\,, \\
\etabar_4^\N &=& C_{\etabar_4 \etabar_2} \left( \etabar_2^N \right)^{9/5}\,, 
\ea
where the coefficients for the $n=1$ polytropes are given by
\allowdisplaybreaks
\ba
C_{\lambdabarnew_3 \lambdabarnew_2}^{(n=1)} &=& -\frac{(2 \pi^2-21 ) \pi^{4/5}}{3^{3/5} (15-\pi^2 )^{7/5}} \approx 0.165\,, \\
C_{\lambdabarnew_4 \lambdabarnew_2}^{(n=1)} &=& \frac{ 3^{4/5} \pi^{8/5} (\pi^4 -105 \pi^2 + 945)}{35 (15-\pi^2 )^{14/5}} \approx 0.0269\,, \\
C_{\sigmabar_2 \lambdabarnew_2}^{(n=1)} &=& -\frac{3^{1/5} (\pi^4 -20 \pi^2+120)}{20 \pi^{8/5} (15-\pi^2 )^{6/5}} \approx -0.0281\,, \\
C_{\etabar_3 \etabar_2}^{(n=1)} &=& \frac{7 \;10^{3/5} \pi^{4/5} (15- \pi^2 )}{2250} \approx 0.159\,, \\
C_{\etabar_4 \etabar_2}^{(n=1)} &=& \frac{3 \pi^{8/5} (21-2 \pi^2)}{35 \; 10^{4/5} (15- \pi^2)} \approx 0.0208\,, 
\ea
and the ones for the $n=0$ polytropes are given by
\allowdisplaybreaks
\ba
C_{\lambdabarnew_3 \lambdabarnew_2}^{(n=0)} &=& \frac{2^{2/5}}{10} \approx 0.132\,, \\
C_{\lambdabarnew_4 \lambdabarnew_2}^{(n=0)} &=& \frac{2^{4/5}}{105} \approx 0.0166\,, \\
C_{\sigmabar_2 \lambdabarnew_2}^{(n=0)} &=& -\frac{2^{1/5}}{70} \approx -0.0164\,, \\
C_{\etabar_3 \etabar_2}^{(n=0)} &=&  \frac{7 \; 3^{2/5}}{2 \; 5^{12/5}} \approx 0.114\,, \\
C_{\etabar_4 \etabar_2}^{(n=0)} &=& \frac{3^{9/5}}{7 \; 5^{14/5}} \approx 0.0114\,.
\ea
These analytic results in the Newtonian limit are shown in Figs.~\ref{fig:lambdabar34-lambdabar2},~\ref{fig:lambdabar2mag-lambdabar2} and~\ref{fig:lambdabar34-lambdabar2-shape} as the dotted-dashed (the $n=1$ polytropes) and dashed (the $n=0$ polytropes) curves respectively. Observe that the numerical values approach such Newtonian curves as one decreases the compactness. Interestingly, the Newtonian relations for the shape tidal deformabilities serve as a good approximation even for high compactness stars. By taking the ratio between the coefficients of the $n=0$ and the $n=1$ polytropes, one finds  
\allowdisplaybreaks
\ba
\frac{C_{\lambdabarnew_3 \lambdabarnew_2}^{(n=0)}}{C_{\lambdabarnew_3 \lambdabarnew_2}^{(n=1)}} &=& \frac{2^{2/5} 3^{3/5} (15-\pi^2)^{7/5}}{10 \pi^{4/5} (21- 2 \pi^2 )} \approx 0.799\,, \\
\frac{C_{\lambdabarnew_4 \lambdabarnew_2}^{(n=0)}}{C_{\lambdabarnew_4 \lambdabarnew_2}^{(n=1)}} &=& \frac{2^{4/5} (15-\pi^2 )^{14/5}}{3^{9/5} \pi^{8/5} (\pi^4 -105 \pi^2 +945)} \approx 0.616\,, \\
\frac{C_{\sigmabar_2 \lambdabarnew_2}^{(n=0)}}{C_{\sigmabar_2 \lambdabarnew_2}^{(n=1)}} &=& \frac{2^{6/5} \pi^{8/5} (15-\pi^2 )^{6/5}}{3^{1/5} 7 (\pi^4 -20 \pi^2 +120)} \approx 0.585\,, \\
\frac{C_{\etabar_3 \etabar_2}^{(n=0)}}{C_{\etabar_3 \etabar_2}^{(n=1)}} &=& \frac{ 3^{12/5} }{2^{3/5} \pi^{4/5} (15- \pi^2 ) } \approx 0.719\,, \\
\frac{C_{\etabar_4 \etabar_2}^{(n=0)}}{C_{\etabar_4 \etabar_2}^{(n=1)}} &=& \frac{ 6^{4/5}(15- \pi^2 ) }{5 \pi^{8/5} (21- 2 \pi^2 ) } \approx 0.547\,.
\ea
If the universality between the $n=1$ and the $n=0$ polytropes in the Newtonian limit holds exactly, the ratio of the coefficients becomes unity. One sees that the deviation from the universality between these two EoSs in the limit is to 20--45\%. Therefore, the universality is not as good as the one for the I-Love-Q relations, where the deviation from the universality between the two EoSs has been shown to be less than 1\%~\cite{I-Love-Q-PRD}. From Figs.~\ref{fig:lambdabar34-lambdabar2},~\ref{fig:lambdabar2mag-lambdabar2} and~\ref{fig:lambdabar34-lambdabar2-shape},  one sees that the universality holds better among the realistic EoSs.

{\renewcommand{\arraystretch}{1.2}
\begin{table*}
\begin{centering}
\begin{tabular}{cccccccccc}
\hline
\hline
\noalign{\smallskip}
$y_i$ & $x_i$ & NS/QS &&  \multicolumn{1}{c}{$a_i$} &  \multicolumn{1}{c}{$b_i$}
&  \multicolumn{1}{c}{$c_i$} &  \multicolumn{1}{c}{$d_i$} &  \multicolumn{1}{c}{$e_i$}  \\
\hline
\noalign{\smallskip}
$\bar{\lambda}_3$ & $\bar{\lambda}_2$ & NS  && -1.15  & 1.18 & $2.51 \times 10^{-2}$  & $-1.31\times 10^{-3}$ & $2.52\times 10^{-5}$\\
 &  & QS  && -0.554  & 0.863 & $7.99 \times 10^{-2}$  & $-5.39\times 10^{-3}$ & $1.35\times 10^{-4}$\\
$\bar{\lambda}_4$ & $\bar{\lambda}_2$ & NS  && -2.45  & 1.43 & $3.95 \times 10^{-2}$  & $-1.81\times 10^{-3}$ & $2.80\times 10^{-5}$\\
& & QS  && -1.33  & 0.808 & 0.146  & $-9.82\times 10^{-3}$ & $2.46\times 10^{-4}$\\
$\bar{\sigma}_2$ & $\bar{\lambda}_2$ & NS  && 0.126  & 0.617 & $2.81 \times 10^{-2}$  & $3.59\times 10^{-4}$ & $-3.61\times 10^{-5}$\\
& & QS  && 0.517  & 0.340 & $7.50 \times 10^{-2}$  & $-3.55\times 10^{-3}$ & $7.21\times 10^{-5}$\\
$\bar{\eta}_3$ & $\bar{\eta}_2$ & NS && -1.64  & 1.36 & $2.86 \times 10^{-3}$  & $2.65\times 10^{-5}$ & $-2.31\times 10^{-6}$\\
 & & QS && 0.494  & 0.514 & 0.119  & -$7.29 \times 10^{-3}$ & $1.67\times 10^{-4}$\\
$\bar{\eta}_4$ & $\bar{\eta}_2$ & NS  && -4.89  & 2.38 & -0.103  & $7.57 \times 10^{-3}$ & $-1.91\times 10^{-4}$\\
&  & QS  && $4.85 \times 10^{-2}$  & 0.318 & 0.198  & $-1.21 \times 10^{-2}$ & $2.77\times 10^{-4}$\\
\noalign{\smallskip}
\hline
\hline
\end{tabular}
\end{centering}
\caption{Estimated numerical coefficients for the fitting formula of the multipole Love relations for NSs and QSs given in Eq.~\eqref{fit}.}
\label{table:coeff}
\end{table*}
}

\section{Applications to Gravitational-Wave Physics}
\label{sec:GWs}

In this section, we show how the universal relations found in this paper bring about benefits to GW parameter estimation. For simplicity, we consider non-spinning NS binaries with masses $m_1$ and $m_2$ in a circular orbit\footnote{For spinning NS binaries with the finite size effect, see e.g.~\cite{I-Love-Q-Science,I-Love-Q-PRD,measuring-EoS}.}.

\subsection{Gravitational Waveform Phase}
\label{sec:phase}

In this subsection, we show how tidal deformabilities enter in the gravitational waveform phase.
Let us denote the gravitational waveform $h(t)$ in the time domain as 
\be
\label{eq:waveform-t}
h(t) = 2 A(t) \cos \phi (t)\,,
\ee
where $A(t)$ and $\phi(t)$ are the amplitude and the phase, respectively. One can express the waveform $\tilde{h}(f)$ in the Fourier domain as 
\be
\label{waveform}
\tilde{h}(f) = \mathcal{A}(f) \exp[i\Psi(f)]\,, 
\ee
where, again, $\mathcal{A}(f)$ and $\Psi(f)$ are the amplitude and phase, respectively. To the leading PN order, the sky-averaged amplitude is given by~\cite{cutlerflanagan}
\be
\mathcal{A}(f) = \frac{1}{\sqrt{30} \pi^{2/3}} \frac{\mathcal{M}^{5/6}}{D_L}f^{-7/6}\,.
\ee
where $D_L$ is the luminosity distance and $\mathcal{M} \equiv m \eta^{3/5}$ is the chirp mass with the total mass $m \equiv m_1 + m_2$ and the symmetric mass ratio $\eta \equiv m_1 m_2/m^2$.
$\Psi(f)$ can be divided into a contribution from the point particle limit $\Psi^{\PP}$ and the one from the finite size effect, $\Psi^{\FS}$. To full order in $\eta$, $\Psi^{\PP} (f)$ is known analytically up to 3.5PN order (up to $n=7$)~\cite{arun35PN}.

Next, let us focus on $\Psi^{\FS} (f)$. 
We first estimate the PN order of the electric-type tidal correction. Two types of tidal corrections to the point-particle phase exist, namely (i) conservative and (ii) dissipative. Let us first look at the former. The radial acceleration in the binary system $a_r^{\bar{\lambda}_\ell}$ due to the tidally-induced $\ell$-th mass multipole moment $M_L^T \propto \bar{\lambda}_\ell \left( \partial_L r^{-1} \right)|_{r=r_{12}}$ is proportional to~\cite{racine} 
\be
a_r^{\bar{\lambda}_\ell} \propto \frac{M_L^T}{r_{12}^{\ell+2}} \propto \frac{\bar{\lambda}_\ell}{r_{12}^{2\ell +3}}\,,
\ee
where $r_{12}$ is the separation between body 1 and 2.  Comparing this to the leading Newtonian point-particle acceleration, $a_r^\PP \propto r_{12}^{-2}$, one finds the conservative electric-type tidal correction as 
\be
\frac{a_r^{\bar{\lambda}_\ell}}{a_r^\PP} \propto \frac{\bar{\lambda}_\ell}{r_{12}^{2 \ell +1}} \propto \bar{\lambda}_\ell v^{2(2 \ell +1)}\,.
\ee
Here, $v$ is the orbital velocity of the binary constituent and we have used the Kepler's Law, $r_{12}^{-1} \propto v^2$. 

On the other hand, one can derive the tidal dissipative correction by calculating corrections to the energy flux.
The energy flux due to the tidally-induced $\ell$-th mass multipole moment is given by 
\ba
\dot{E}^{\bar{\lambda}_\ell} &\propto & \frac{d^{\ell +1} M_L^B}{d t^{\ell+1}} \frac{d^{\ell +1} M_L^T}{d t^{\ell+1}} \nn \\ 
&\propto & \bar{\lambda}_\ell \omega^{2(\ell+1)} \frac{r_{12}^\ell}{r_{12}^{\ell + 1}} \propto \bar{\lambda}_\ell v^{2(3 \ell +4)}\,,
\ea
where $\omega (\propto v^3)$ is the orbital angular velocity and $M_L^B \propto r_{12}^\ell$ is the $\ell$-th binary mass multipole moment. Comparing this to the leading quadrupole point-particle emission~\cite{cutlerflanagan}, $\dot{E}^\PP \propto v^{10}$, one finds 
\be
\frac{\dot{E}^{\bar{\lambda}_\ell}}{\dot{E}^\PP} \propto \bar{\lambda}_\ell v^{2 (3 \ell -1)}\,.
\ee
Therefore, the conservative and dissipative tidal corrections become comparable in the PN sense for $\ell=2$, and the former dominates the latter for $\ell > 2$. The electric-type tidal correction is of $(2 \ell +1)$PN order relative to the leading Newtonian.

Next, we estimate the PN order of the magnetic-type tidal correction. For non-spinning binaries, the magnetic-type tidal correction to the radial acceleration does not exist~\cite{racine}. On the other hand, the magnetic-type tidal correction to the energy flux is given by 
\ba
\dot{E}^{\bar{\sigma}_\ell} &\propto & \frac{d^{\ell +1} S_L^B}{d t^{\ell+1}} \frac{d^{\ell +1} S_L^T}{d t^{\ell+1}} \nn \\
& \propto & \bar{\sigma}_\ell \omega^{2(\ell+1)} v^2 \frac{r_{12}^\ell}{r_{12}^{\ell +1}} \propto \bar{\sigma}_\ell v^{2(3 \ell +5)}\,,
\ea
where $S_L^B \propto r_{12}^\ell v$ and $S_L^T \propto \bar{\sigma}_\ell v \left( \partial_L r^{-1} \right)|_{r=r_{12}}$ are the $\ell$-th binary and tidally-induced current multipole moments, respectively. Again, by comparing this to $\dot{E}^\PP$, one finds 
\be
\frac{\dot{E}^{\bar{\sigma}_\ell}}{\dot{E}^\PP} \propto \bar{\sigma}_\ell v^{6 \ell}\,.
\ee
Therefore, the magnetic-type tidal correction is of $(3\ell)$PN order relative to the leading Newtonian. 

The leading finite size effect enters at 5PN order relative to the Newtonian term due to a contribution from $\bar{\lambda}_2$~\cite{flanagan-hinderer-love}, while the one from $\bar{\lambda}_3$ enters at 7PN order~\cite{flanagan-hinderer-love, hinderer-lackey-lang-read}. Extending these results, we derive the leading finite size effect from $\bar{\lambda}_\ell$ on the gravitational waveform phase as 
\ba
\label{psi-lambda-l}
\Psi_{\bar{\lambda}_\ell} &=& - \sum_{A=1}^2 \left[ \frac{5}{16} \frac{(2\ell -1)!! (4\ell +3) (\ell +1)}{(4\ell -3) (2\ell -3)}  \bar{\lambda}_{\ell}^A X_A^{2\ell -1} x^{2 \ell -3/2} \right. \nn \\
& & \left. +  \frac{9}{16} \delta_{\ell 2}  \bar{\lambda}_{2}^A \frac{X_A^4}{\eta} x^{5/2} \right] + \mathcal{O} \left( x^{2\ell -1/2} \right)\,,
\ea
where $x\equiv (\pi m f)^{2/3}$, $A,B=1,2$ with $A \neq B$, and $X_A = m_A/m$. (See Appendix~\ref{app:lambda-l} for the derivation.) The first and second terms correspond to the conservative and dissipative contributions respectively. We have checked that the above formula agrees with the previous results for the $\ell=2$ and $\ell=3$ cases mentioned above. The leading 6PN effect due to the contribution from $\bar{\sigma}_2$ is given by
\be
\label{psi-sigma-2}
\Psi_{\bar{\sigma}_2} = \sum_{A=1}^2 \frac{5}{224}  \bar{\sigma}_2^A \frac{X_A^5}{\eta} ( X_A- X_B) x^{7/2} + \mathcal{O}(x^{9/2})\,.
\ee
(See Appendix~\ref{app:sigma_2} for the derivation.) Notice that the leading term in $\Psi_{\bar{\sigma}_2}$ vanishes for an equal-mass binary ($X_1=X_2$). Therefore, the total finite-size effect $\Psi^{\FS} (f)$ can be written as
\ba
\label{phase-FS}
\Psi^{\FS} (f) &=& \sum_{\ell =2} \left( \Psi_{\bar{\lambda}_\ell} + \Psi_{\bar{\sigma}_\ell} \right) \nn \\
 & = & \sum_{\ell=2} \sum_{A=1}^{2}  \left[   \sum_{n=2(2\ell +1)} \psi_{n/2}^{\bar{\lambda}_\ell, A} x^{(-5+n)/2} \right. \nn \\
& & \left. + \sum_{n=6\ell} \psi_{n/2}^{\bar{\sigma}_\ell, A} x^{(-5+n)/2}  \right]\,. \nn \\
\ea
The first six $\psi_{n/2}^{\bar{\lambda}_2, A}$~\cite{damour-nagar-villain}
are listed in Appendix~\ref{app:phase}, while $\psi_{2\ell + 1}^{\bar{\lambda}_\ell, A}$ and $\psi_{3\ell}^{\bar{\sigma}_2, A}$ can be read off from Eqs.~\eqref{psi-lambda-l} and~\eqref{psi-sigma-2} respectively.

\subsection{Useful Number of GW Cycles}
\label{sec:useful}

\begin{figure}[htb]
\begin{center}
\includegraphics[width=8.5cm,clip=true]{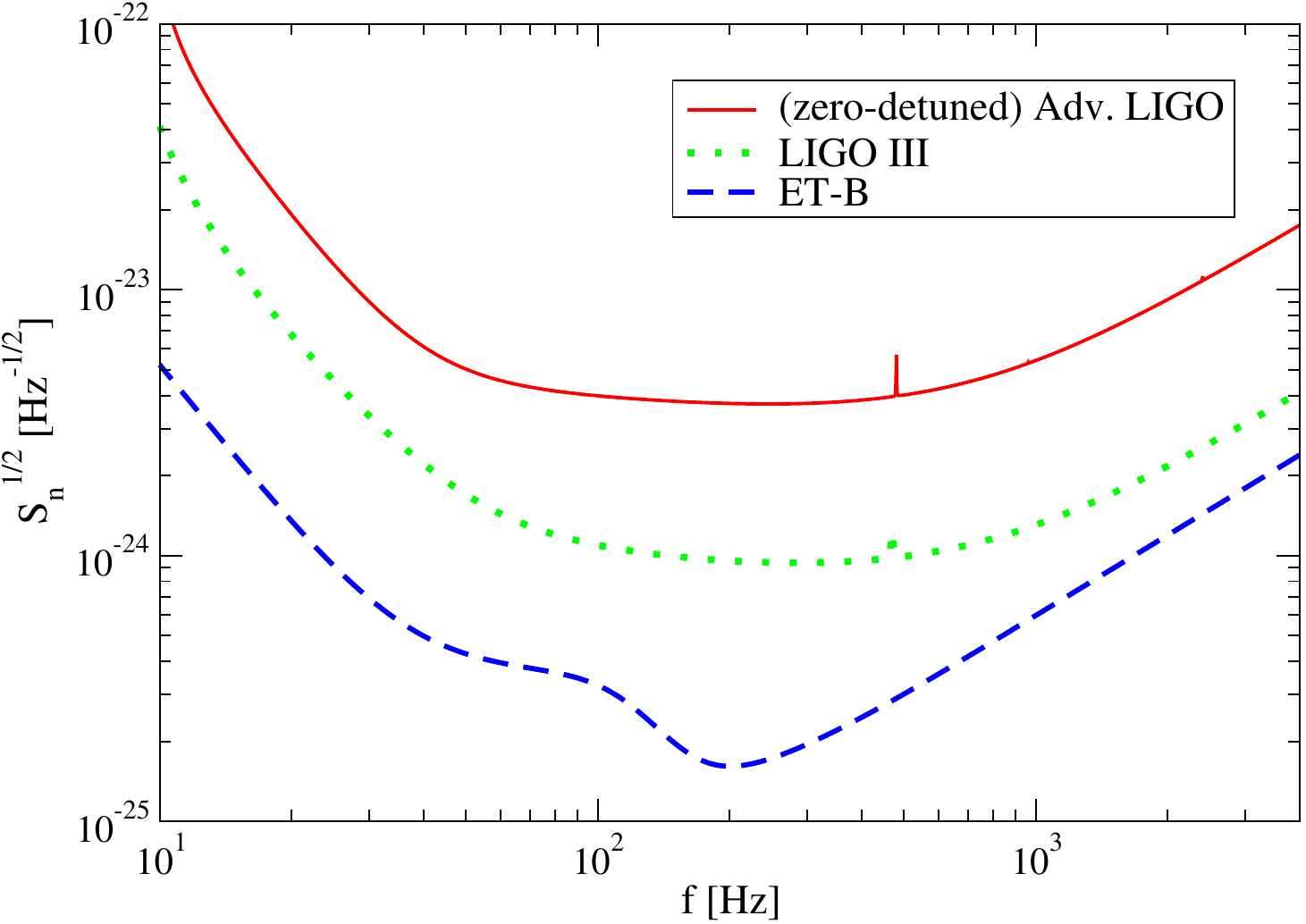}  
\caption{\label{fig:noise} 
(Color online) Noise spectral density for (zero-detuned) Adv. LIGO~\cite{AdvLIGO-noise} (red solid), LIGO III~\cite{LIGO3-noise} (green dotted) and ET-B~\cite{ET-noise} (blue dashed). 
}
\end{center}
\end{figure}

\begin{figure}[htb]
\begin{center}
\includegraphics[width=8.5cm,clip=true]{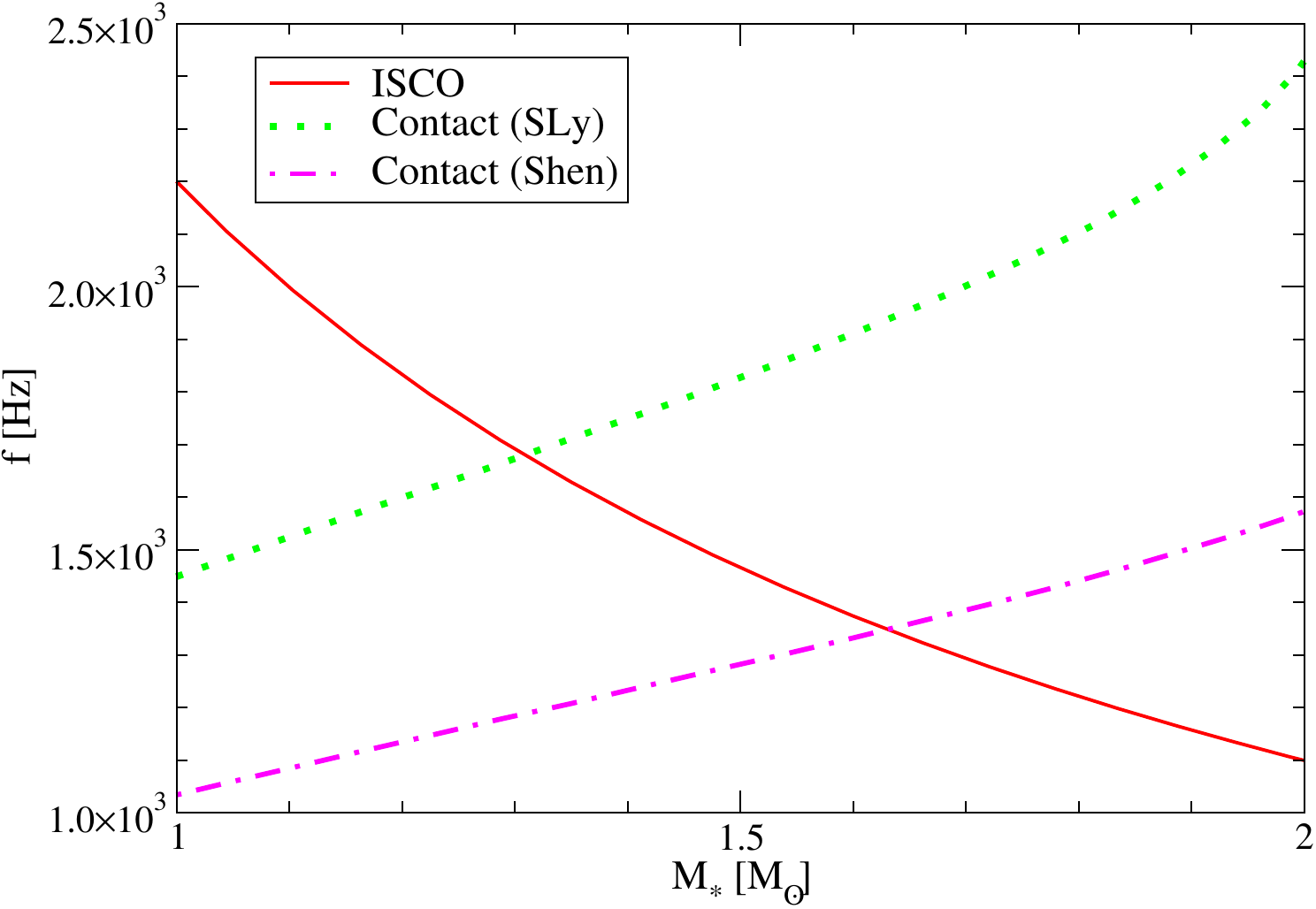}  
\caption{\label{fig:fmax} 
(Color online) Frequency at ISCO (red solid) and contact with SLy (green dotted) and Shen (magenta dotted-dashed) EoSs. We assume equal-mass, non-spinning NS/NS binaries.  
}
\end{center}
\end{figure}

Let us now consider the useful number of GW cycles~\cite{damour-useful} of each term in the phase. 
The optimal SNR $\rho$ is defined by
\be
\rho \equiv \sqrt{(h|h)}\,,
\ee
where the inner product $(A|B)$ is defined by
\be
\label{eq:inner-product}
(A|B) \equiv 2 \int^{f_\mrm{max}}_{f_\mrm{min}} \frac{\tilde{A}^*(f) \tilde{B}(f) + \tilde{A}(f) \tilde{B}^*(f)}{S_n(f)}  df\,,
\ee
with $\tilde{A}$ and $\tilde{B}$ corresponding to the Fourier transform of $A$ and $B$ respectively. 
$S_n(f)$ is the spectral noise density shown in Fig.~\ref{fig:noise} for (zero-detuned) Adv. LIGO~\cite{AdvLIGO-noise}, LIGO III~\cite{LIGO3-noise} and ET-B~\cite{ET-noise}. The fitting formulas of the noise curve for Adv.~LIGO and ET-B are given in Ref.~\cite{ajith-zero-detuned} and Refs.~\cite{mishra,schutz-review} respectively. The one for LIGO III is provided in Appendix~\ref{app:LIGOIII}. $f_\mrm{min}$ and $f_\mrm{max}$ in the above equation are the minimum and maximum cutoff frequencies respectively. We set $f_\mrm{min}=10$ Hz for Adv.~LIGO and LIGO III while $f_\mrm{min}=1$ Hz for ET-B, and $f_\mrm{max} = \min (f_\mrm{ISCO}, f_\mrm{cont})$. Here, $f_\mrm{ISCO} \equiv (6^{3/2} \pi m)^{-1}$ is the frequency at the innermost-stable circular orbit (ISCO) of a point-particle in Schwarzschild spacetime and $f_\mrm{cont}$ is the approximate frequency at contact given by $f_\mrm{cont} = \sqrt{m/(R_{*,1} + R_{*,2})^3}/\pi$. $f_\mrm{cont}$ for the SLy and Shen EoSs, together with $f_\mrm{ISCO}$ are shown in Fig.~\ref{fig:fmax}.

The total number of GW cycles $N$ is defined by~\cite{berti-buonanno}
\be
N = \int^{f_\mrm{max}}_{f_\mrm{min}}  N_\mrm{inst}(f) \ d\ln f\,,
\ee
where the instantaneous number of GW cycles $N_\mrm{inst}(f)$ is defined by
\be
N_\mrm{inst} (f) \equiv \frac{f}{2 \pi} \frac{d \phi}{d f} = \frac{f^2}{\dot{f}}\,. 
\ee
Here we recall that $\phi (t)$ is the GW phase defined in Eq.~\eqref{eq:waveform-t} and we used $\dot{\phi} = 2 \pi f$. $\dot{f}$ is calculated from $\Psi(f)$ as $\dot{f} = (2/\pi) (d^2 \Psi/d \omega^2)^{-1}$ [see Eq.~\eqref{phase-relation}]. However, a more useful quantity would be the one that accounts for the weight of the detector noise. In other words, the only cycles that would be useful are those that contribute most to the SNR. Following~\cite{damour-useful}, we define the useful number of GW cycles $N_\mrm{useful}$ as  
\be
\label{eq:useful}
N_\mrm{useful} = \left( \int^{f_\mrm{max}}_{f_\mrm{min}} \frac{df}{f} w(f) N_\mrm{inst}(f) \right) \left( \int^{f_\mrm{max}}_{f_\mrm{min}} \frac{df}{f} w(f)  \right)^{-1}\,,
\ee
where $w(f)$ is defined as
\be
w(f) \equiv \frac{A(t)^2}{f S_n(f)}\,.
\ee
We remind the reader that $A(t)$ is the GW amplitude defined in Eq.~\eqref{eq:waveform-t}. $t$ is given in terms of $f$ to leading PN order as~\cite{cutlerflanagan} $t(f) = t_c - 5 \mathcal{M}^{-5/3} (8 \pi f)^{-8/3}$ with $t_c$ representing the time of coalescence.

\begin{figure*}[htb]
\begin{center}
\includegraphics[width=8.5cm,clip=true]{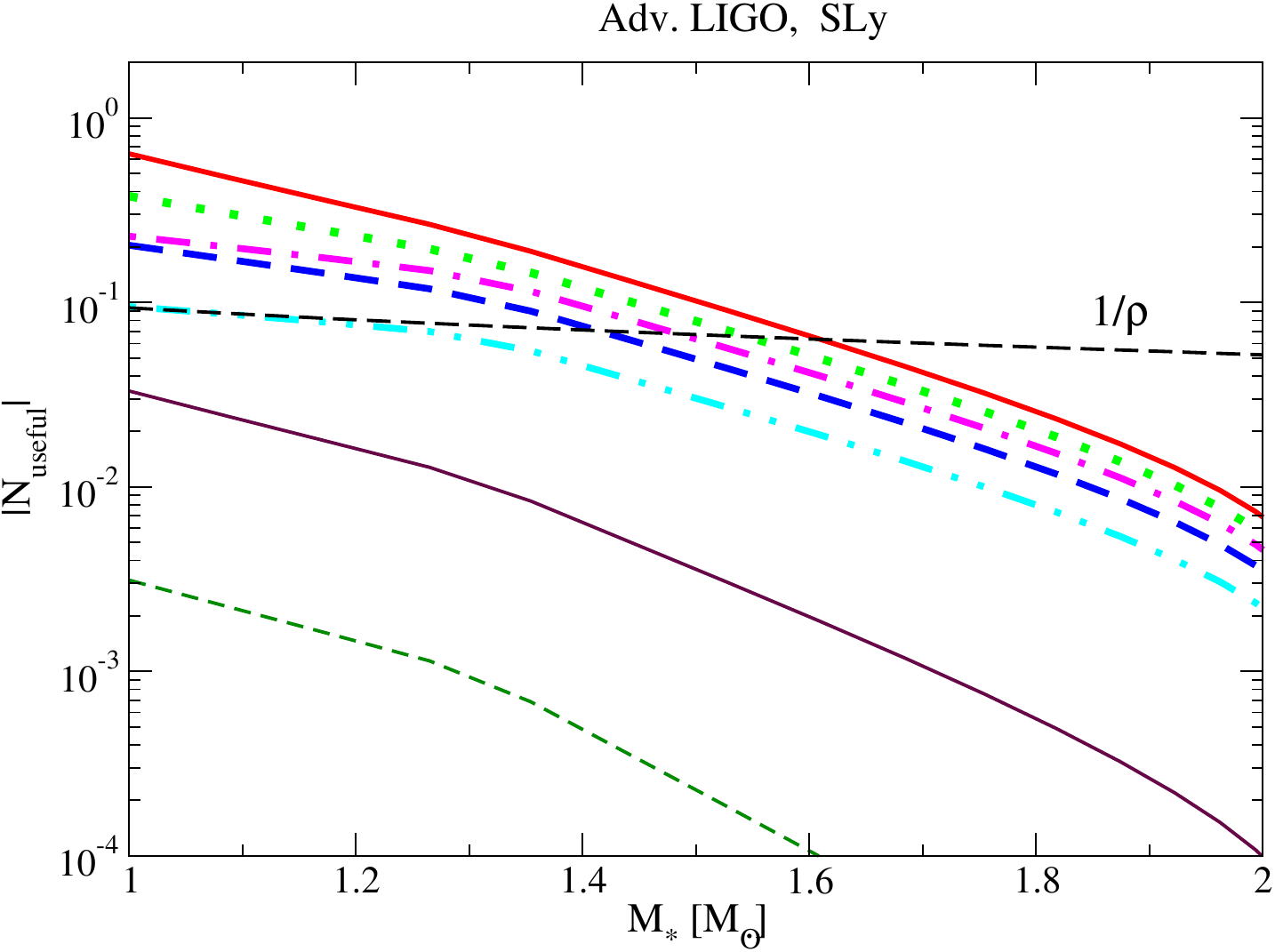}  
\includegraphics[width=8.5cm,clip=true]{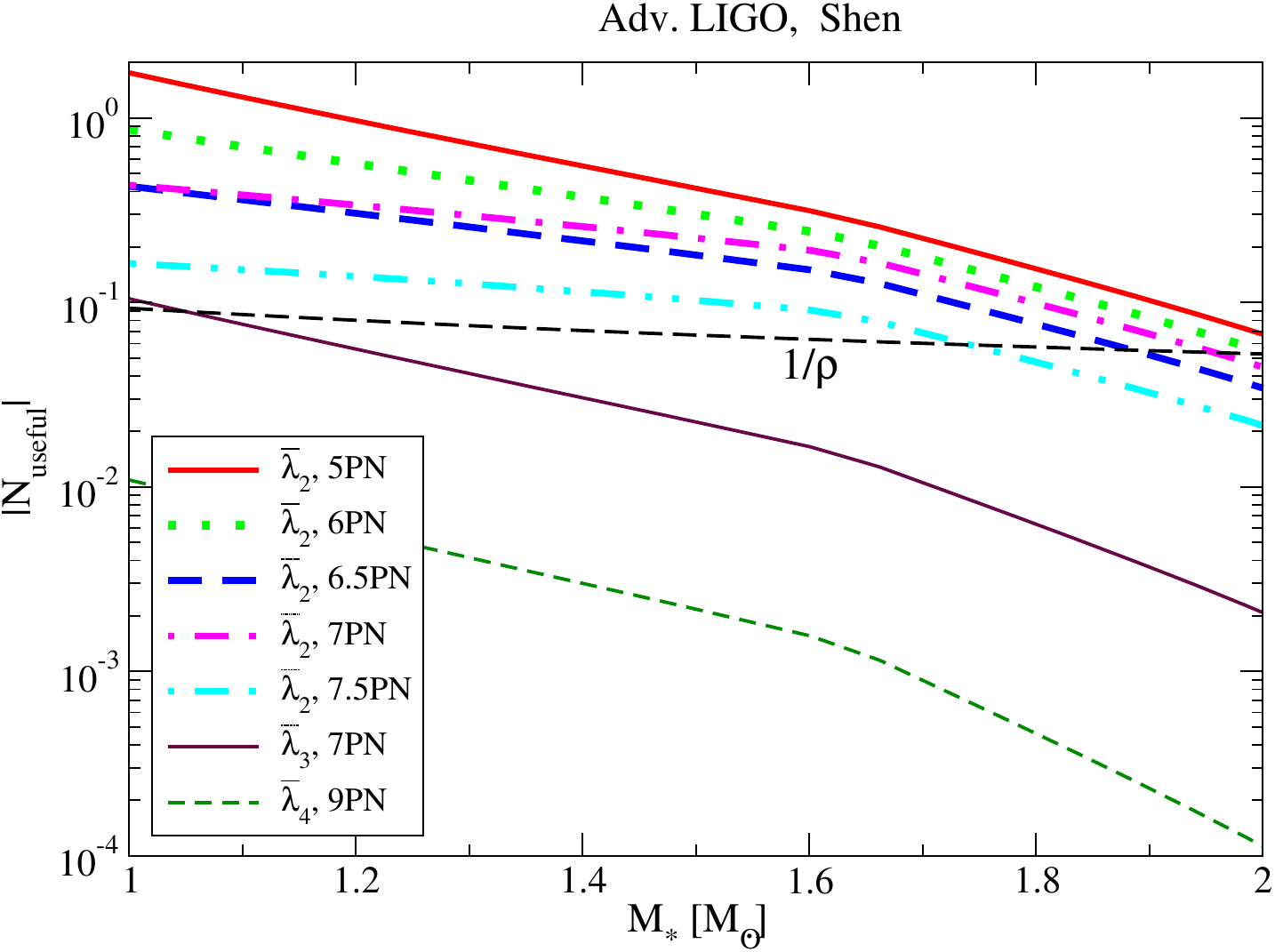}  
\caption{\label{fig:useful-LIGO} 
(Color online) Useful number of cycles for various terms in the GW phase with the SLy (left) and Shen (right) EoSs and using Adv.~LIGO. As a reference, we plot $1/\rho$ for an equal-mass NS binary at $D_L=100$Mpc as a black dashed line. One can roughly say that the useful number of GW cycles above this curve can be important for a parameter estimation. 
}
\end{center}
\end{figure*}

\begin{figure*}[htb]
\begin{center}
\includegraphics[width=8.5cm,clip=true]{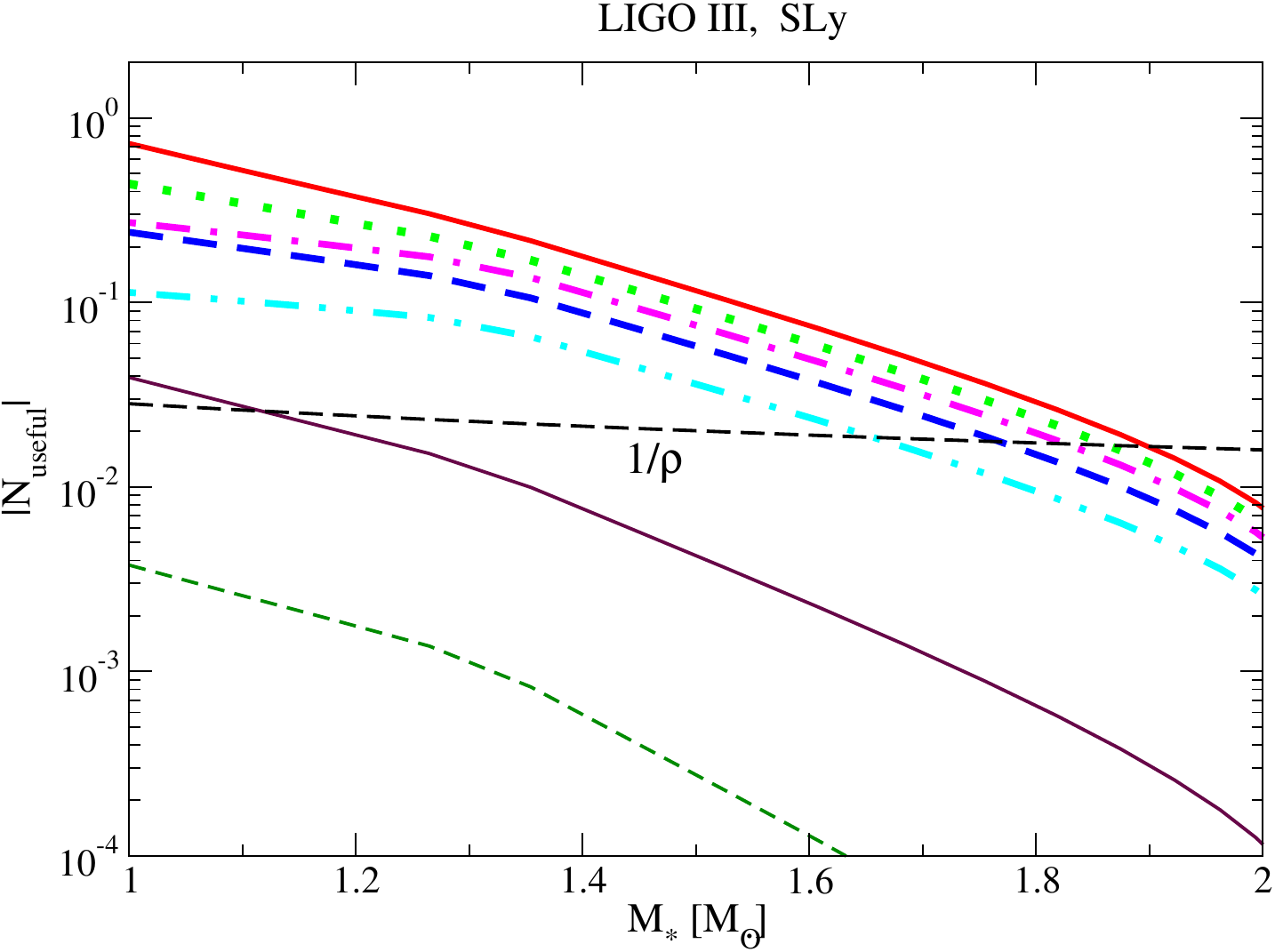}  
\includegraphics[width=8.5cm,clip=true]{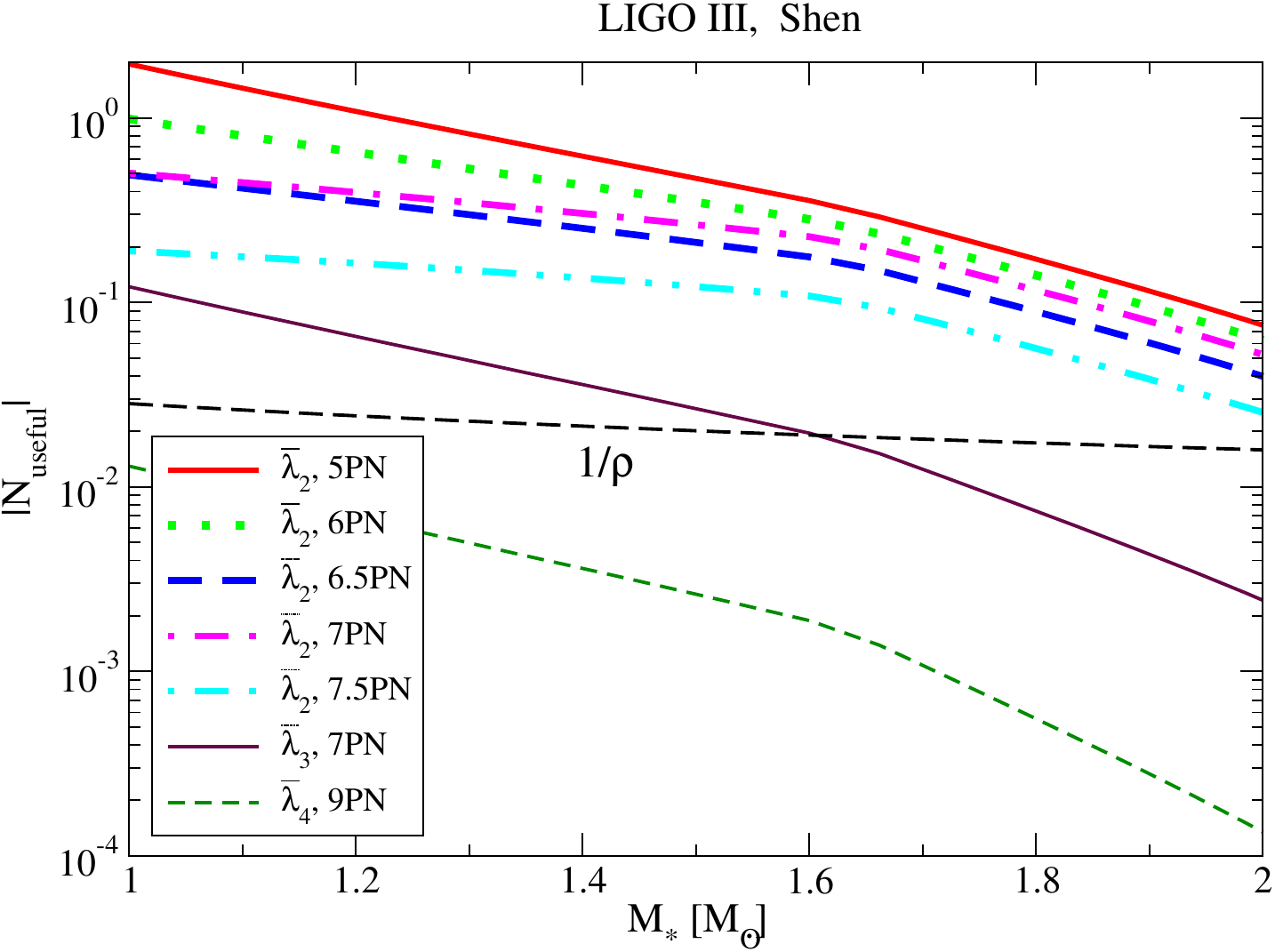}  
\caption{\label{fig:useful-LIGO3} 
(Color online) Same as Fig.~\ref{fig:useful-LIGO} but with LIGO III. 
}
\end{center}
\end{figure*}

\begin{figure*}[htb]
\begin{center}
\includegraphics[width=8.5cm,clip=true]{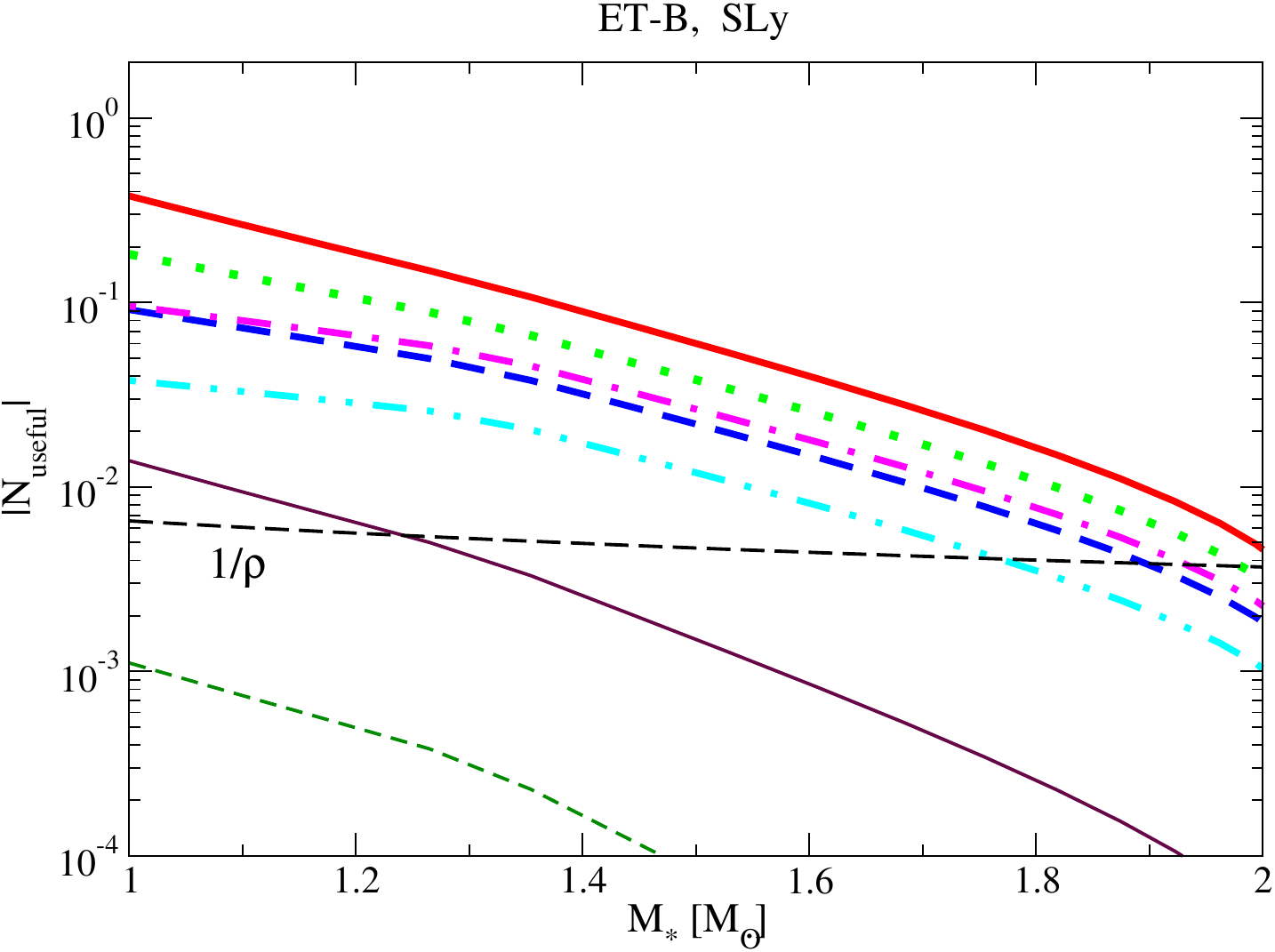}  
\includegraphics[width=8.5cm,clip=true]{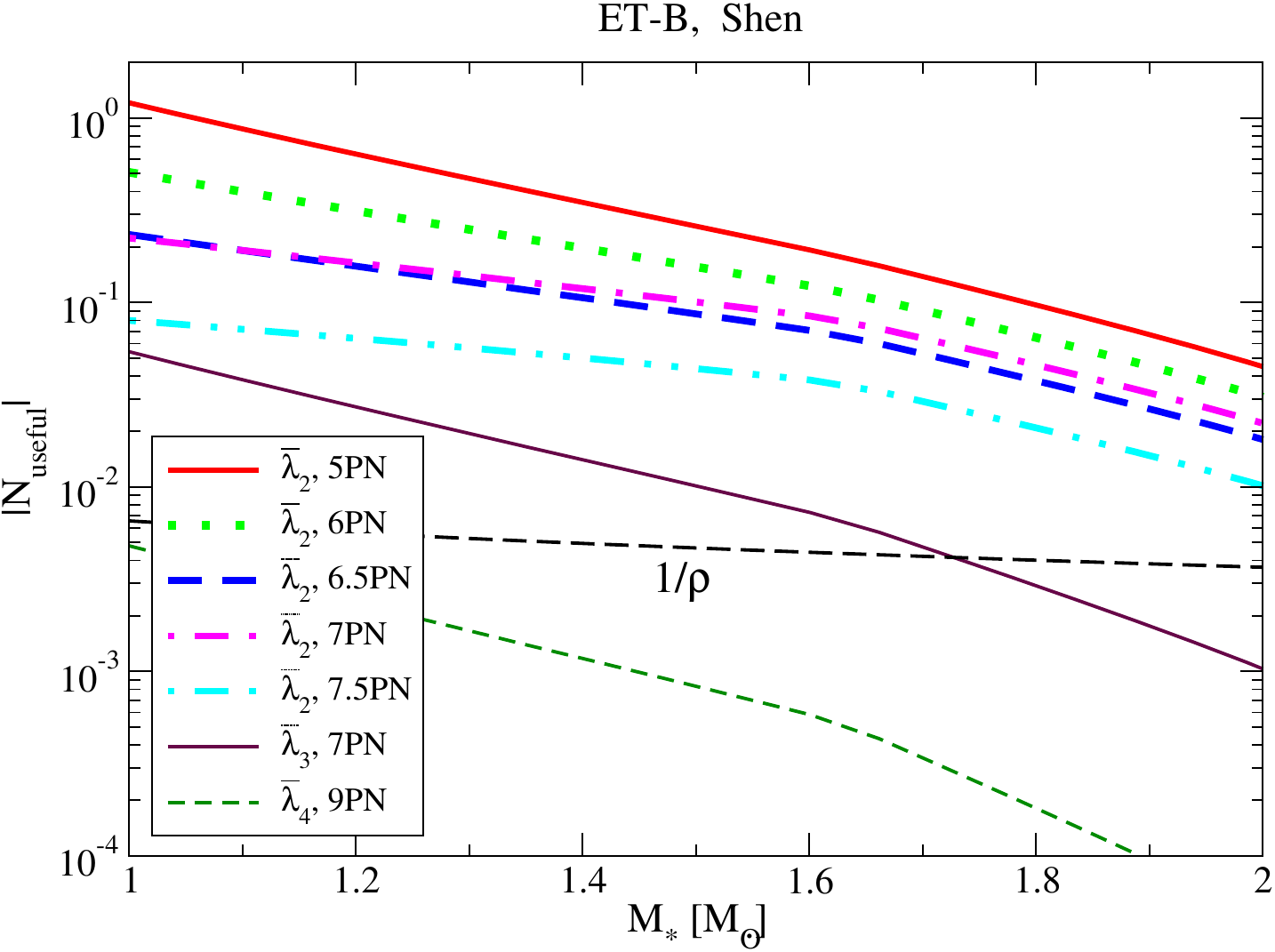}  
\caption{\label{fig:useful-ET} 
(Color online) Same as Fig.~\ref{fig:useful-LIGO} but with ET-B. 
}
\end{center}
\end{figure*}

Figures~\ref{fig:useful-LIGO},~\ref{fig:useful-LIGO3} and~\ref{fig:useful-ET} show useful number of GW cycles of each finite-size term in the GW phase against a NS mass for the SLy (left) and Shen (right) EoSs with Adv~LIGO, LIGO-III and ET-B. We assumed equal-mass, non-spinning NS/NS binaries when calculating the plot. We have checked that the useful number of GW cycles for the APR and LS220 EoSs lie in between the SLy and Shen ones shown in the plots. The contribution from $\sigmabar_2$ is not shown in the plot since the leading contribution of $\sigmabar_2$ in GW phase vanishes for an equal-mass binary.
For a reference, we show $1/\rho$ at the luminosity distance of $D_L = 100$ Mpc with black dashed curves. Roughly speaking, any terms in $\dot{f}$ [or equivalently, $\Psi(f)$] whose useful number of GW cycles are above the $1/\rho$ curve would affect GW parameter estimation. Since the useful number of GW cycles are normalized as in Eq.~\eqref{eq:useful}, it reflects the shape of the noise curve but not the overall magnitude. Figures~\ref{fig:useful-LIGO} and~\ref{fig:useful-LIGO3} show that the useful number of GW cycles for Adv.~LIGO and LIGO-III are similar, since the shape of the noise curves is similar. On the other hand, the ET-B curves in Fig.~\ref{fig:useful-ET} have values smaller than those for Adv.~LIGO and LIGO-III. This is because the latter shape is broader, and hence are more sensitive on the high-frequency region.


Now, let us focus on $N_\mrm{useful}$ from the tidal contribution. One also sees that $N_\mrm{useful}$ decreases as one increases the NS mass. This is because the NS compactness decreases as one increases the NS mass, which then decreases $\lambdabarnew_\ell$. Figures~\ref{fig:useful-LIGO},~\ref{fig:useful-LIGO3} and~\ref{fig:useful-ET} show that $\lambdabarnew_3$ is important in performing parameter estimation for LIGO III and ET-B. In this case, the universal $\lambdabarnew_3$--$\lambdabarnew_2$ relation helps in expressing $\lambdabarnew_3$ in terms of $\lambdabarnew_2$, which reduces the number of parameters and breaks the degeneracy between the tidal parameters.

\subsection{Parameter Estimation}

We will next show the importance and the impact of using the universal relations found in this paper on a parameter estimation. 
The statistical error of parameters $\theta^i$ can be estimated by~\cite{cutlerflanagan,finn,vallisneri-fisher} 
\be
\label{eq:stat}
\Delta \theta^i_\stat = \sqrt{\left( \Gamma^{-1} \right)_{ii}}\,,
\ee
where $\Gamma_{ij}$ is the Fisher matrix defined by
\be
\Gamma_{ij} \equiv ( \partial_i h | \partial_j h)\,,
\ee
with $\partial_i \equiv \partial/\partial \theta^i$ and the inner product is define by Eq.~\eqref{eq:inner-product}.

As we explained in Sec.~\ref{sec:phase}, the GW phase can be decomposed into the point-particle and finite-size parts.
For the former, we keep up to 3.5PN order. Higher PN terms can give non-negligible systematic errors as shown in~\cite{favata-sys,kent-sys}, but we here focus on the statistical errors on tidal deformabilities and have checked that the inclusion of the such terms to leading order in the mass ratio does not affect the statistical errors. For the latter phase,  we keep the first six terms (up to 7.5PN) that depend on $\lambdabarnew_2$ and the leading term of $\lambdabarnew_3$. We define the averaged dimensionless $\ell$-th electric tidal deformability $\lambdabarnew_{\ell,s}$ and its difference $\lambdabarnew_{\ell,a}$ as~\cite{I-Love-Q-PRD}
\be
\label{lambdabar_s}
\lambdabarnew_{\ell,s} \equiv \frac{\lambdabarnew_{\ell}^1 + \lambdabarnew_{\ell}^2}{2}, \quad \lambdabarnew_{\ell,a} \equiv \frac{\lambdabarnew_{\ell}^1 - \lambdabarnew_{\ell}^2}{2}\,,
\ee
where $\lambdabarnew_{\ell}^A$ denotes $\lambdabarnew_\ell$ of the $A$-th body.
Notice that $\lambdabarnew_{\ell,s} = \lambdabarnew_{\ell}^1 =\lambdabarnew_{\ell}^2$ and $\lambdabarnew_{\ell,a} = 0$ for an equal-mass binary.
Then, the terms that depend on $\lambdabarnew_{2}$ and $\lambdabarnew_{3}$ in the phase can be expressed as 
\ba
\Psi_{\lambdabarnew_2} &=& - \frac{9}{16} \frac{x^{5/2}}{\eta} \left[ \lambdabarnew_{2,s}  \hat{\Psi}_{\lambdabarnew_{2,s}}(X_1,X_2,x)  \right. \nn \\
& & \left.  +  \lambdabarnew_{2,a}  \hat{\Psi}_{\lambdabarnew_{2,a}}(X_1,X_2,x) \right]\,, \\
\Psi_{\lambdabarnew_3} &=& - \frac{125}{12} x^{9/2} \left[ \lambdabarnew_{3,s}  \hat{\Psi}_{\lambdabarnew_{3,s}}(X_1,X_2,x)  \right. \nn \\
& & \left.  +  \lambdabarnew_{3,a}  \hat{\Psi}_{\lambdabarnew_{3,a}}(X_1,X_2,x) \right]\,, 
\ea
where $\hat{\Psi}_{\lambdabarnew_{2,s}}$, $\hat{\Psi}_{\lambdabarnew_{2,a}}$, $\hat{\Psi}_{\lambdabarnew_{3,s}}$ and $\hat{\Psi}_{\lambdabarnew_{3,a}}$ are functions of $X_1$, $X_2$ and $x$. 
Following~\cite{damour-nagar-villain}, for simplicity, we neglect the mass dependence on $\hat{\Psi}_{\lambdabarnew_{2,s}}$ and $\hat{\Psi}_{\lambdabarnew_{3,s}}$ when calculating the Fisher matrix.

In this subsection, we assume equal-mass NS binaries with a circular orbit and neglect the effect of $\lambdabarnew_{\ell,a}$. Therefore, we have seven parameters in total:
\be
\theta^i = (\ln \mathcal{M}, \ln \eta, t_c, \phi_c, \ln D_L, \lambdabarnew_{2,s}, \lambdabarnew_{3,s})\,.
\ee
For fiducial signal values, we set $\eta = 1/4$, $t_c = 0 = \phi_c$, $D_L = 100$Mpc and vary $M_* (= m_1= m_2)$.

Figure~\ref{fig:delta-lambda2-LIGO3} shows the statistical errors on $\ln \lambdabarnew_{2,s}$ against $M_*$ with and without using the universal relations for LIGO III and ET-B with the SLy and Shen EoSs. Since $\lambdabarnew_{\ell,s} = \lambdabarnew_{\ell}^A$ for an equal-mass binary, one can use the $\lambdabarnew_3$--$\lambdabarnew_2$ relation found in the previous section as the $\lambdabarnew_{3,s}$--$\lambdabarnew_{2,s}$ relation. When calculating the one with the universal relations, we do not include $\lambdabarnew_{3,s}$ into parameters since one can express them in terms of $\lambdabarnew_{2,s}$. For the parameter estimation without using the universal relations, we include $\lambdabarnew_{3,s}$ into parameters. 
One sees that the measurement accuracy increases by a factor of 5 if one uses the relation, essentially reducing the statistical errors to the level where $\lambdabarnew_{3,s}$ is not taken into account\footnote{Statistical errors with the universal relation are not the same as the ones where $\lambdabarnew_{3,s}$ is not taken into account since the former has an additional information on $\lambdabarnew_{2,s}$ from the term in the GW phase proportional to $\lambdabarnew_{3,s}$. However, we have checked that the difference is rather small.}. 
In the figure, we also show systematic errors on $\ln \lambdabarnew_{2,s}$ due to the difference between the fitting formula and actual values on $\lambdabarnew_{3,s}$, shown in the bottom left panel of Fig.~\ref{fig:lambdabar3-lambdabar2} or Fig.~\ref{fig:lambdabar34-lambdabar2}. One sees that statistical errors dominate systematic errors due to the fitting, which validates the use of the fitting formula for the universal relations in GW parameter estimation.


\subsection{Unequal-Mass Systems}

Now, let us relax the assumption that $m_1 = m_2$ and consider GWs from unequal-mass binaries. For such binaries, in principle, one needs to account for both $\lambdabarnew_{2,s}$ and $\lambdabarnew_{2,a}$ in the model tidal parameters. However, this will produce a strong correlation between these two tidal parameters and will deteriorate the measurement accuracy of $\lambdabarnew_{2,s}$ compared to the equal-mass case. Therefore, we first ask the following question: Up to what mass difference can one safely neglect the contribution of $\lambdabarnew_{2,a}$? 

\begin{figure}[htb]
\begin{center}
\includegraphics[width=8.5cm,clip=true]{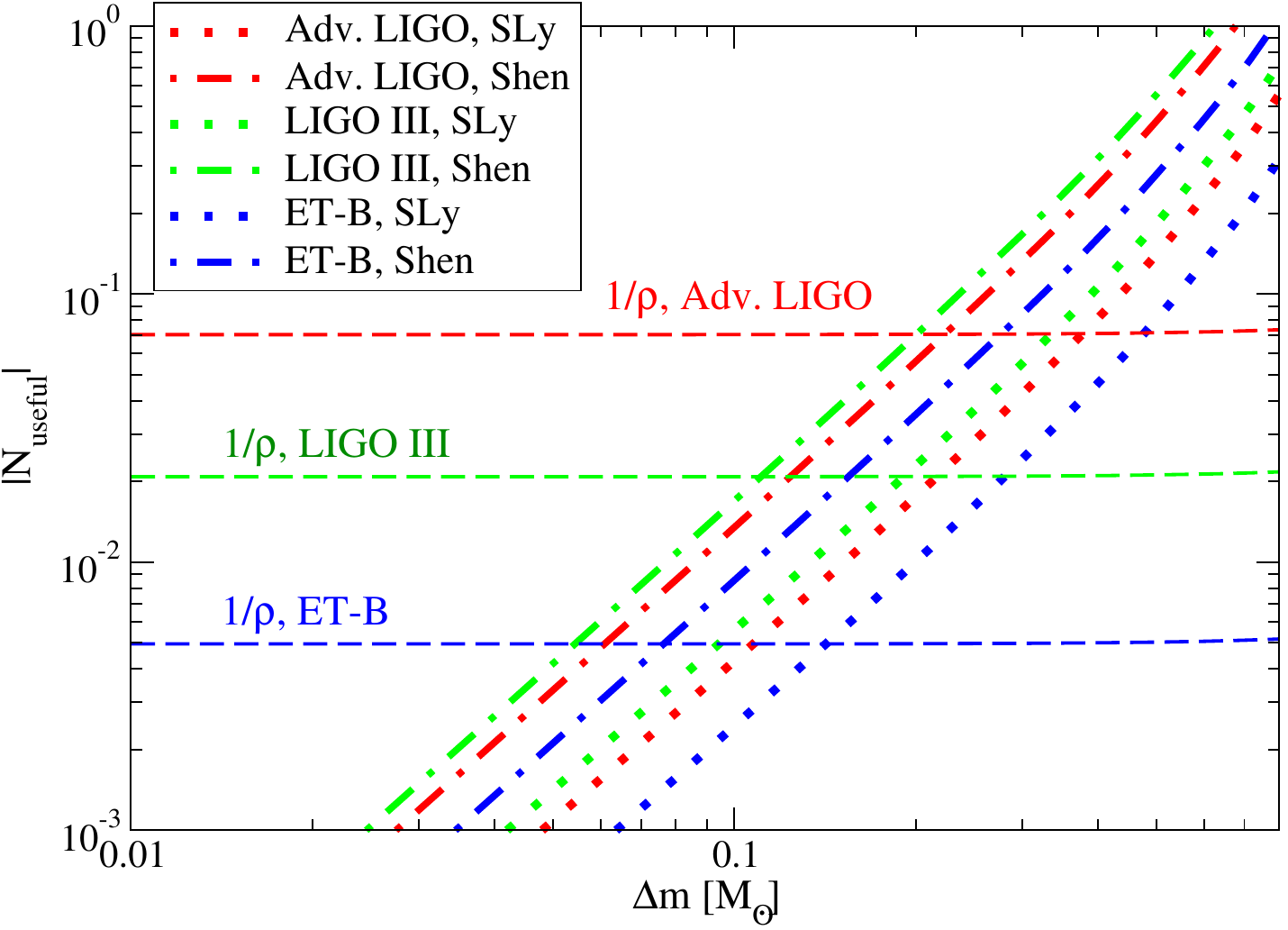}  
\caption{\label{fig:useful-unequal} 
(Color online) Useful number of GW cycles for terms proportional to $\lambdabarnew_{2,a}$ in the GW phase against $\Delta m \equiv m_1 - m_2$, where we set the averaged mass to $m/2 = 1.4 M_\odot$. We consider using Adv.~LIGO (red), LIGO III (green) and ET-B (blue) with the SLy (dotted) and Shen (dotted-dashed) EoSs. As a reference, we plot $1/\rho$ for a NS binary at $D_L=100$Mpc for each all detectors. Observe that $N_\mrm{useful}$ is below the threshold $1/\rho$ if $\Delta m < 0.2M_\odot$, $\Delta m < 0.1 M_\odot$ and $\Delta m < 0.07 M_\odot$ for Adv.~LIGO, LIGO III and ET-B respectively. 
}
\end{center}
\end{figure}

We can answer this question by calculating the useful number of GW cycles of terms that are proportional to $\lambdabarnew_{2,a}$ in the GW phase, which we present in Fig.~\ref{fig:useful-unequal}. Here, we set the averaged mass to $m/2 = 1.4 M_\odot$ and plot $N_\mrm{useful}$ against $\Delta m = m_1 - m_2$ for Adv.~LIGO (red), LIGO III (green) and ET-B (blue) with the SLy (dotted) and Shen (dotted-dashed) EoSs. For reference, we also plot $1/\rho$ for each detector, assuming that the source is at $D_L=100$Mpc. Again, $\lambdabarnew_{2,a}$ would affect parameter estimation if $N_\mrm{useful}$ is above the threshold $1/\rho$. One sees that $\lambdabarnew_{2,a}$ has a larger effect with the Shen EoS than the SLy one (if one fixes $\Delta m$). Figure~\ref{fig:useful-unequal} shows that for Adv.~LIGO, one does not need to take $\lambdabarnew_{2,a}$ into account if $\Delta m < 0.2 M_\odot$ for $m/2 = 1.4M_\odot$. For LIGO III and ET-B, the allowed range on $\Delta m$ reduces to $\Delta m < 0.1 M_\odot$ and $\Delta m < 0.07 M_\odot$ respectively.

\begin{figure}[htb]
\begin{center}
\includegraphics[width=8.5cm,clip=true]{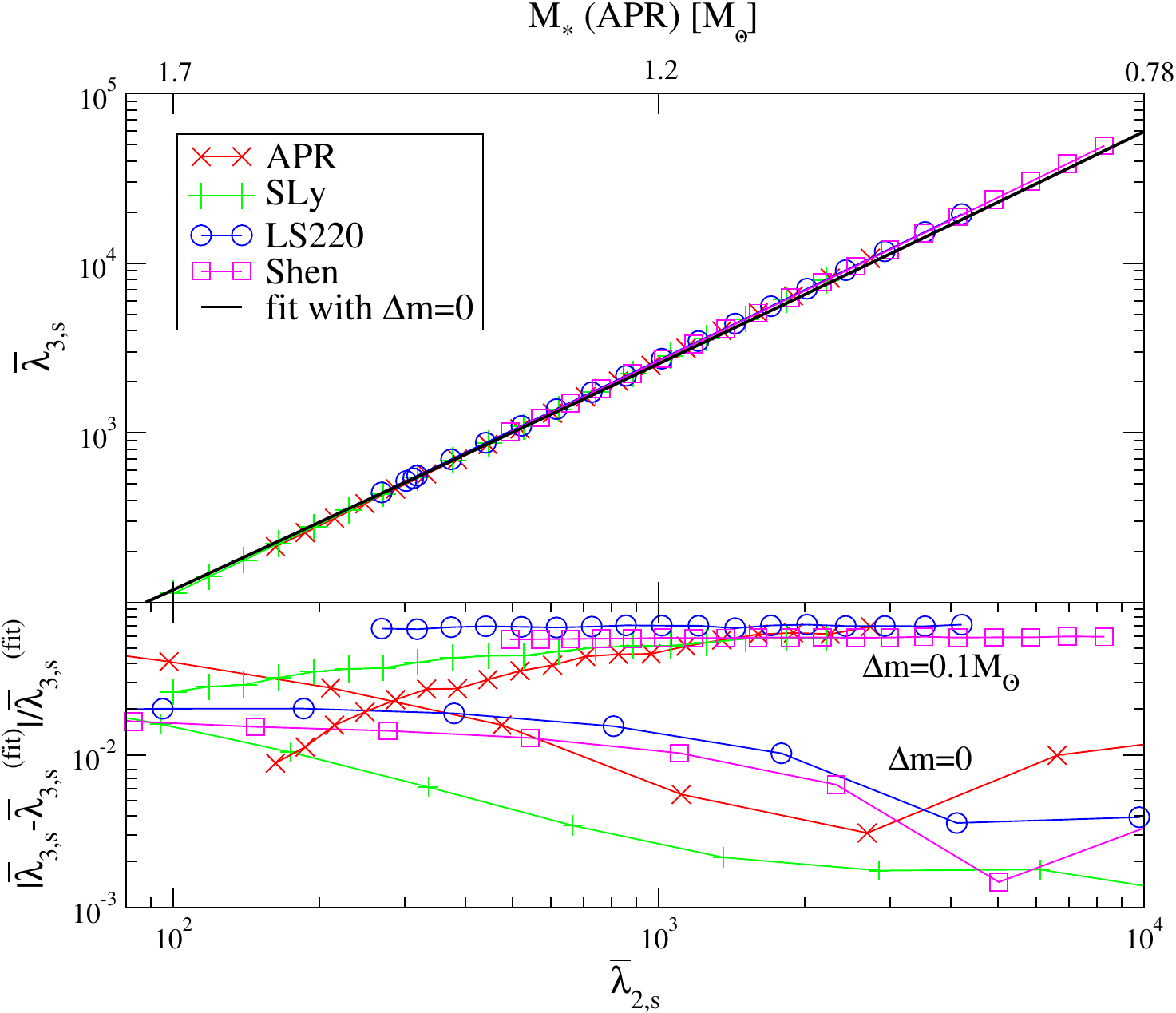}  
\caption{\label{fig:lambdabar3-lambdabar2-unequal} 
(Color online) (Top) The universal $\lambdabarnew_{3,s}$--$\lambdabarnew_{2,s}$ relation for both equal-mass (which is the same as the $\lambdabarnew_{3}$--$\lambdabarnew_{2}$ relation shown in Figs.~\ref{fig:lambdabar3-lambdabar2} and~\ref{fig:lambdabar34-lambdabar2}) and unequal-mass systems with $\Delta m = 0.1 M_\odot$. We also show the fit (black solid) obtained for $\Delta m =0$.
(Bottom) Fractional difference between the numerical values and the fit for $\Delta m =0$.
 Observe that the the fit for $\Delta m =0$ still holds to $\mathcal{O}(5)$\% accuracy for $\Delta m = 0.1 M_\odot$.
}
\end{center}
\end{figure}

For an unequal-mass system, one might wonder if one is still allowed to use the universal $\lambdabarnew_{3}$--$\lambdabarnew_{2}$ relation found in this paper that reduces to the universal $\lambdabarnew_{3,s}$--$\lambdabarnew_{2,s}$ relation in the equal-mass case. The top panel of Fig.~\ref{fig:lambdabar3-lambdabar2-unequal} presents the $\lambdabarnew_{3,s}$--$\lambdabarnew_{2,s}$ relation for both $\Delta m = 0$ and $\Delta m = 0.1M_\odot$. For reference, we show the fit with the black solid curve obtained for equal-mass binaries. On the bottom panel, we show the fractional difference between each curve and the fit. One sees that although the fractional difference becomes larger for unequal-mass binaries, the fit for equal-mass binaries is still valid to $\mathcal{O}(5)$\% accuracy in the parameter range shown in the figure, even when $\Delta m = 0.1M_\odot$. This shows that for binaries where  $\lambdabarnew_{2,a}$ can safely be neglected, one can still use the $\lambdabarnew_{3}$--$\lambdabarnew_{2}$ relation as the $\lambdabarnew_{3,s}$--$\lambdabarnew_{2,s}$ one.

Finally, let us here discuss how $\sigmabar_2$ would contribute to {\emph {nearly}} equal-mass binaries. We first take the ratio of $\psi_6^{\sigmabar_2}$ and $\psi_6^{\lambdabarnew_2}$ (which both give a 6PN term contribution to the phase relative to the point-particle Newtonian), which for $m_1 \approx m_2$ becomes
\ba
\frac{\psi_6^{\sigmabar_2}}{\psi_6^{\lambdabarnew_2}} & \approx & -\frac{16}{13083} (X_1-X_2) \frac{\sigmabar_2}{\lambdabarnew_2} \nn \\
& \approx & \mathcal{O}(10^{-5}) \left( \frac{\delta_m}{0.03} \right)\,,
\ea
where $\delta_m \equiv X_1-X_2 [=(m_1-m_2)/m]$ and we used $|\sigmabar_2| \sim \lambdabarnew_2$, as can be seen in Fig.~\ref{fig:lambdabar2mag-lambdabar2}.
This shows that for nearly equal-mass binaries, the leading $\sigmabar_2$ contribution on $N_\mrm{useful}$ is $\mathcal{O}(10^5)$ times smaller than the contribution from the 6PN $\lambdabarnew_2$ contribution. Therefore, one can safely neglect the contribution from $\sigmabar_2$ when performing parameter estimation.

\section{Conclusions and Discussions}
\label{sec:conclusions}

Similar to the I-Love-Q relations~\cite{I-Love-Q-Science,I-Love-Q-PRD}, we found new universal relations among the electric, magnetic and shape dimensionless tidal deformabilities of NSs and QSs that do not depend strongly on the EoS. Especially, the QS tidal deformabilities (except for the $\ell=2$ electric tidal one~\cite{hinderer-lackey-lang-read,postnikov}) have been calculated for the first time. 

Such universal relations are especially useful to GW physics since they allow us to reduce the number of tidal parameters that are needed to be fitted via matched filtering. First, we derive the leading contribution of $\lambdabarnew_\ell$ and $\sigmabar_2$ to the GW phase from a NS binary inspiral and found that they enter at $2 \ell +1$ and $6$ PN orders, respectively. 
We next calculated the useful number of GW cycles of each finite-size terms in the phase and showed that, for LIGO III and ET-B, one needs to take $\lambdabarnew_{3,s}$ into account as well as $\lambdabarnew_{2,s}$ for parameter estimation. As recommended in~\cite{damour-nagar-villain}, it would be better to include such higher multipolar tidal deformabilities when analyzing real GW data.

We then estimated the statistical errors on $\ln \lambdabarnew_{2,s}$ using a Fisher analysis, where we include both $\lambdabarnew_{2,s}$ and $\lambdabarnew_{3,s}$ into the model tidal parameters. By using the universal $\lambdabarnew_3$--$\lambdabarnew_2$ relation, one can break the degeneracy between $\lambdabarnew_{2,s}$ and $\lambdabarnew_{3,s}$, and improves the measurement accuracy of $\ln \lambdabarnew_{2,s}$ by a factor of 5, which essentially reduces to the one where only $\lambdabarnew_{2,s}$ is included into the model tidal parameters. The use of the universal relation is limited by the systematic error due to the fit. We have checked that the statistical errors on $\ln \lambdabarnew_{2,s}$ are larger than the systematic error from the fit. However, such systematic errors can become larger just before merger, where the adiabatic approximation breaks down and one needs to consider dynamical tides instead~\cite{maselli}.

Since the tidal deformability parameters are different for the two binary constituents, in principle, one needs to account for two $\ell=2$ electric tidal parameters. In this paper, we used $\lambdabarnew_{2,s}$ and $\lambdabarnew_{2,a}$ to parameterize this dependence. We estimated the useful number of GW cycles for $\lambdabarnew_{2,a}$ and, comparing this to the $1/\rho$ threshold for a source at $D_L=100$Mpc, we found that one can safely neglect $\lambdabarnew_{2,a}$ for GW parameter estimation with Adv.~LIGO if $\Delta m < 0.2M_\odot$ and an averaged mass $m/2 = 1.4M_\odot$. We also showed that the $\lambdabarnew_{3,s}$--$\lambdabarnew_{2,s}$ relation can still be valid for unequal-mass binaries to $\mathcal{O}(5)$\% accuracy for $\Delta m < 0.1M_\odot$.

Although $\lambdabarnew_{2,s}$ and $\lambdabarnew_{2,a}$ have a clear physical meaning, a better choice of parameters may exist for parameter estimation. One possible choice is the one made in~\cite{favata-sys}, where one takes the primary tidal parameter to be the one that enters at 5PN order (and such parameter agrees with the individual tidal parameter for an equal-mass binary). Then, unlike $\lambdabarnew_{2,a}$ that also enters at 5PN order, the other parameter (which vanishes for an equal-mass binary) enters only at 6PN order. This means that the systematic error due to not including this second parameter for nearly equal-mass system becomes smaller than that discussed in this paper.

We checked that the statistical errors on $\ln \lambdabarnew_{3,s}$ with Adv.~LIGO, LIGO III and ET-B are all above unity, which means that the detectors considered here are not sensitive enough to actually measure $\ln \lambdabarnew_{3,s}$. This is because a strong correlation exists between $\lambdabarnew_2$ and $\lambdabarnew_3$. However, they can still place an upper bound, which allows us to perform a model-independent and EoS-independent test of GR, similar to the one explained in~\cite{I-Love-Q-Science,I-Love-Q-PRD}. Namely, one can draw an error box around a measured central value in the $\lambdabarnew_2$--$\lambdabarnew_3$ plane. One can rule out any coupling constants in modified gravity theories that do not produce the curve in this plane that goes through the error box. Such a possibility is worth looking at in future, together with the calculation of the tidal deformabilities in theories other than GR~\cite{I-Love-Q-Science,I-Love-Q-PRD,quadratic}.

In this paper, we focused on GWs from non-spinning NS/NS binary inspirals in quasi-circular orbits. As discussed in~\cite{favata-sys}, the systematic errors on $\ln \lambdabarnew_{2,s}$ due to not including the NS's spins and eccentricities are negligible compared to the statistical errors. On the other hand, the systematic errors due to not including higher PN terms can be important, as showed in~\cite{favata-sys,kent-sys}.

Although we found that $\lambdabarnew_{3,s}$ can be important for parameter estimation with third-generation GW detectors, other unknown tidal terms in the GW phase can also be important. For example, the 7PN tidal term has not been calculated completely yet. One may expect that the unknown 7PN tidal terms will be smaller than the known lower-order terms~\cite{damour-nagar-villain}. However, this does not mean that such unknown terms will be unimportant in GW parameter estimation. The useful number of GW cycles of such unknown terms can be larger than those of $\lambdabarnew_{3,s}$. It is important to estimate the unknown tidal terms in the GW phase that depend on $\lambdabarnew_{2}$ to sufficiently high PN order.

A parameter estimation study for spinning NS binaries, using not only the Fisher but also a Bayesian analyses, shall be discussed elsewhere~\cite{measuring-EoS}. It is important to estimate tidal deformability parameters for spinning NSs. Although the linear tidal parameters defined in Eq.~\eqref{eq:linear-tidal} are the most generic ones for non-spinning NSs, one needs to introduce other tidal parameters for spinning NSs that would mix the electric and magnetic contributions~\cite{damour-nagar}. Similarly, it would be interesting to investigate how magnetic fields would affect tidal deformabilities. 
One might be able to tackle these problems perturbatively, assuming either the spins or magnetic fields are small.

\section*{Acknowledgments}

The author thanks Nicol\'as Yunes for valuable discussions and comments, and reading the manuscript carefully. The author also thanks Leo Stein for helping him carry out some of the STF calculations. The author thanks Tanja Hinderer for checking some of the calculations in~\cite{hinderer-lackey-lang-read}.
K.Y. acknowledges support from NSF grant PHY-1114374, NSF CAREER Grant PHY-1250636 and NASA grant NNX11AI49G. Some calculations used the computer algebra-systems MAPLE, in combination with the GRTENSORII package~\cite{grtensor}.

\appendix

\section{Derivation of the Tidal Gravitational Waveform Phase}
\label{app:phase2}

In this appendix, we show the derivation of $\Psi_{\bar{\lambda}_\ell} (f)$ and $\Psi_{\bar{\sigma}_2} (f)$[Eqs.~\eqref{psi-lambda-l} and~\eqref{psi-sigma-2} respectively]. For simplicity, we assume that the 1st body is tidally-deformed due to the 2nd one and treat the latter as a point-particle. The final expression in the main text can be obtained by changing the indices 1 and 2, and linearly combining the 2 pieces together.

\subsection{The $\ell$-th Electric Tidal Deformability}
\label{app:lambda-l}

We start with finding the conservative tidal correction to the binding energy. The radial acceleration $a_r^{\lambdabarnew_\ell^1}$ due to the tidally-induced $\ell$-th multipole moment of the 1st body $M_{L,1}^T$ is given by~\cite{racine}
\ba
\label{ar-lambdal}
a_r^{\lambdabarnew_\ell^1} &=& \frac{m}{r_{12}^2} \left[ 1+ \frac{(-1)^\ell  (2 \ell +1)!!}{\ell !} \frac{M_{L,1}^{T} n_{\langle r L \rangle}}{m_1 r_{12}^\ell} \right] \nn \\
&=& \frac{m}{r_{12}^2} \left[ 1 + (2 \ell-1)!! (\ell+1) \lambdabarnew_{\ell}^1 \eta X_1^{2 \ell -1} \left(\frac{m}{r_{12}} \right)^{2\ell + 1} \right]\,, \nn \\
\ea 
where $n^i = (x_1^i - x_2^i)/r_{12}$ with $x^i_A$ denoting the position of the $A$-th body and $X_A = m_A/m$ for $A=1,2$. In the second equality, we have used
\ba
M_{L,1}^T & = & - \lambda_{\ell}^1 \partial_L \left( -\frac{m_2}{r} \right) \Big|_{r=r_{12}} \nn \\
& =&  (-1)^{\ell} (2 \ell -1)!! \; \lambdabarnew_{\ell}^1 m_1^{2 \ell +1} m_2 \frac{n_L}{r_{12}^{\ell +1}}\,,
\ea
and the STF identity
\be
n_{\langle i L \rangle} n_{L} = \frac{(\ell +1)!}{(2 \ell +1)!!} n_i\,,
\ee
which can be obtained from the identities~\cite{blanchet-damour}
\ba
n_{\langle i L \rangle} &=& n_i n_{L} - \frac{\ell}{2 \ell +1} \delta_{i \langle a_1} n_{a_2 \cdots a_\ell \rangle}\,, \\ 
n_L n_L &=& \frac{\ell !}{(2 \ell -1)!!} \,,
\ea
and solving a recurrence equation.

 Equating Eq.~\eqref{ar-lambdal} with $r_{12} \omega^2$, solving for $r_{12} (\omega)$ and expanding in $v = (m \omega)^{1/3} \ll 1$, one obtains the tidally-modified Kepler's Law as
\be
r_{12} = \left( \frac{m}{\omega^2} \right)^{1/3} \left[ 1 + \frac{(2 \ell-1)!! (\ell+1)}{3} \lambdabarnew_{\ell}^1 \eta X_1^{2 \ell-1} x^{2 \ell+1} \right]\,,
\ee
where we recall $x = v^2 =(m \omega)^{2/3} = (\pi m f)^{2/3}$. Integrating the equation of motion with respect to $r_{12}$, one obtains the binding energy $E(r_{12})$. Then, substituting the modified Kepler's Law above, one finds
\be
\label{E-lambda-l}
E = - \frac{1}{2} \eta m x \left[ 1  -\frac{(2 \ell -1)!! (4 \ell + 1)}{3} \lambdabarnew_{\ell}^1  \eta X_1^{2 \ell-1}  x^{2 \ell+1} \right]\,.
\ee

Next, we look at the dissipative tidal correction. As we discuss in Sec.~\ref{sec:phase}, the dissipative tidal correction is always subdominant for $\ell >2$, and hence, to leading order, only $\ell=2$ matters.  Substituting $M_{\langle ij \rangle} = \eta m r_{12}^2 n_{\langle ij \rangle} + \delta_{\ell 2} M_ {\langle ij \rangle, 1}^T$ into the quadrupolar radiation energy flux formula, $\dot{E} = - \langle \dddot{M}_{\langle ij \rangle} \dddot{M}_{\langle ij \rangle} \rangle /5$~\cite{kidder}, where $\langle \cdots \rangle$ denotes orbital averaging, one finds
\ba
\label{Edot-lambda-l}
\dot{E} &=& - \frac{32}{5} \eta^2 x^{5} \left[ 1 + \frac{4 (2 \ell -1)!! (\ell +1)}{3} \lambdabarnew_{\ell}^1 \eta X_1^{2 \ell-1} x^{2 \ell+1}  \right. \nn \\
& & \left. +  6 \delta_{\ell 2} \lambdabarnew_{2}^1 X_1^{4} x^5 \right]\,,
\ea
where we used the modified Kepler's Law.

With these conservative and dissipative tidal corrections at hand, one can calculate the correction to the gravitational waveform phase in the Fourier domain from the relation~\cite{tichy}
\be
\label{phase-relation}
\frac{d^2 \Psi}{d \omega^2} = 2 \frac{dE}{d\omega} \frac{dt}{dE}\,.
\ee
Plugging Eqs.~\eqref{E-lambda-l} and~\eqref{Edot-lambda-l} into the above formula and integrating with respect to $\omega$ twice, one finds
\ba
\Psi (f) &=& \frac{3}{128} \frac{1}{\eta} x^{-5/2} \left[ 1 \right. \nn \\
& & \left. - \frac{40}{3} \frac{(2\ell -1)!! (4\ell +3) (\ell +1)}{(4\ell -3) (2\ell -3)} \bar{\lambda}_{\ell}^1 \eta X_1^{2\ell -1}  x^{2 \ell +1} \right. \nn \\
& & \left. - 24 \delta_{\ell 2} \bar{\lambda}_{2}^1 X_1^4  x^5 \right]\,.
\ea
One recovers Eq.~\eqref{psi-lambda-l} by taking the correction part of the above equation, changing the index of body 1 to body 2, and linearly combining the two parts together.

\subsection{The $\ell=2$ Magnetic Tidal Deformability}
\label{app:sigma_2}

One can obtain $\Psi_{\sigmabar_2}(f)$ in a similar manner to Sec.~\ref{app:lambda-l}. As we discuss in Sec.~\ref{sec:phase}, there is no magnetic-type tidal conservative correction for non-spinning binaries. For the dissipative correction, one needs the total current quadrupole moment, which is given by $S_{\langle ij \rangle} = - [\mu \delta_m r_{12}^2 - 12 \mu \sigmabar_{2}^1/r_{12}^3] \epsilon_{kl( j} n_{i) k} v_l$. Here, $\delta_m \equiv (m_1 - m_2)/m$ and $v^i \equiv v_1^i - v_2^i$. The first term corresponds to the binary current quadrupole moment~\cite{kidder} while the second term denotes the tidally-induced current quadrupole moment.  Substituting this and $M_{\langle ij \rangle} = \eta m r_{12}^2 n_{\langle ij \rangle}$ into the radiation energy flux formula, $\dot{E} = - [\langle \dddot{M}_{\langle ij \rangle} \dddot{M}_{\langle ij \rangle} + (16/9) \dddot{S}_{\langle ij \rangle} \dddot{S}_{\langle ij \rangle} \rangle ]/5$~\cite{kidder}, one finds
\be
\label{Edot-sigma-2}
\dot{E} = - \frac{32}{5} \eta^2 x^{5} \left[ 1 - \frac{2}{3} \sigmabar_{2}^1 (X_1 - X_2) X_1^5 x^6 \right]\,,
\ee

Substituting $E = -  \eta m x/2$ and Eq.~\eqref{Edot-sigma-2} into Eq.~\eqref{phase-relation} and integrating with respect to $\omega$ twice, one obtains $\Psi (f)$ as
\be
\Psi (f) = \frac{3}{128} \frac{1}{\eta} x^{-5/2} \left[ 1 + \frac{20}{21} \sigmabar_{2}^1 (X_1 - X_2) X_1^5  x^6 \right]\,.
\ee
Again, one recovers Eq.~\eqref{psi-sigma-2} by taking the tidal correction part of the above equation, exchanging the indices 1 and 2, and combining the two pieces together. Notice that $\Psi_{\sigmabar_2}(f) = 0$ for an equal-mass binary.

{\renewcommand{\arraystretch}{1.2}
\begin{table*}
\begin{centering}
\begin{tabular}{cccccc}
\hline
\hline
\noalign{\smallskip}
$c_0$ & $c_1$ & $c_2$ &  $c_3$ &  $c_4$   \\
\hline
\noalign{\smallskip}
-3.029396206 & 0.2842933209 & -149.1949551 & 315.0564131  & -318.3482493\\
\noalign{\smallskip}
\hline
\noalign{\smallskip}
$c_5$ & $c_6$ & $c_7$ & $c_8$ & $c_9$\\
\hline
\noalign{\smallskip}
187.6402670  & -67.83575993 & 14.84129658 &-1.806200458 & 0.09393770700\\
\noalign{\smallskip}
\hline
\hline
\end{tabular}
\end{centering}
\caption{Estimated numerical coefficients for the fitting formula of the LIGO III noise curve given by Eq.~\eqref{eq:fit-LIGOIII}.}
\label{table:coeff-LIGOIII}
\end{table*}
}

\section{Various Coefficients in Gravitational Waveform Phase}
\label{app:phase}

In this appendix, we list some of the coefficients in the gravitational waveform phase given in Eq.~\eqref{phase-FS}. The first six $\psi_{n/2}^{\bar{\lambda}_2, A}$~\cite{flanagan-hinderer-love,vines2,hinderer-lackey-lang-read,damour-nagar-villain} are given by
\allowdisplaybreaks
\ba
\psi_5^{\bar{\lambda}_2, A} &=& -\frac{9}{16} \frac{X_A^5}{\eta} \left( 1 + 12 \frac{X_B}{X_A} 
\right) \bar{\lambda}_2^{A}\,, \\
\psi_{5.5}^{\bar{\lambda}_2, A} &=& 0\,, \\
\psi_6^{\bar{\lambda}_2, A} &=& \frac{5 (3179-919X_A-2286X_A^2+260X_A^3)}{672(12-11X_A)} 
\psi_{5}^{\bar{\lambda}_2, A}\,, \nn \\
\\
\psi_{6.5}^{\bar{\lambda}_2, A} &=& -\pi \psi_5^{\bar{\lambda}_2, A}\,, \\
\psi_{7}^{\bar{\lambda}_2, A} &=& \frac{1}{12-11X_A} \left( \frac{39927845}{508032} - \frac{480043345}{9144576}X_A \right. \nn \\
& & \left. + \frac{9860575}{127008}X_A^2 - \frac{421821905}{2286144}X_A^3 + \frac{4359700}{35721}X_A^4 \right. \nn \\
& & \left. - \frac{10578445}{285768} X_A^5 \right) \psi_5^{\bar{\lambda}_2, A} + \cdots\,,\\
\psi_{7.5}^{\bar{\lambda}_2, A} &=& -\frac{\pi}{672 (12-11 X_A)} \left( 27719-22127 X_A \right. \nn \\
& & \left. +7022 X_A^2-10232 X_A^3 \right) \psi_5^{\bar{\lambda}_2, A}\,, 
\ea
\begin{figure}[htb]
\begin{center}
\includegraphics[width=8.5cm,clip=true]{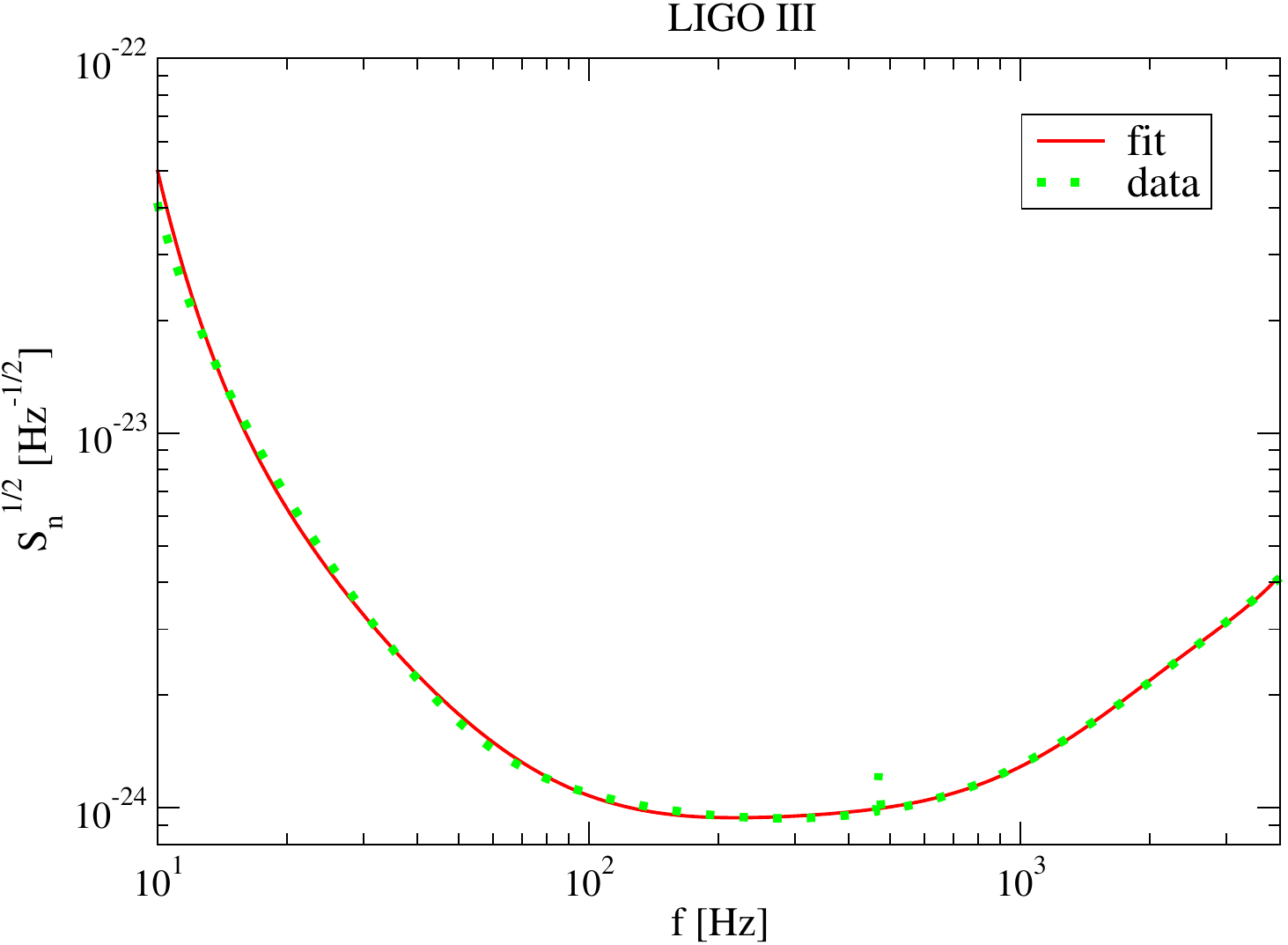}  
\caption{\label{fig:noise-LIGOIII} 
(Color online) Noise spectral density for LIGO III~\cite{LIGO3-noise}. We compare the fit (red solid) with the data (green dotted).
 }
\end{center}
\end{figure}
where $X_A = m_A/m$.
$\psi_{n/2}^{\bar{\lambda}_2, A}$ shown above are complete up to $n=13$~\cite{damour-nagar-villain}. At $n=14$, some missing terms (denoted by ``$\cdots$'') exist but the order of magnitude estimate shows that they are subdominant~\cite{damour-nagar-villain}. $n=15$ contribution comes from the tail effect~\cite{damour-nagar-villain}.

In~\cite{hinderer-lackey-lang-read}, it reads that the tidal effect proportional to $(\lambdabarnew_2)^2$ enters at 7.5PN order in the gravitational wave phase, but this is a typo and such an effect enters at 10PN order. The terms proportional to $(\lambdabarnew_2)^2$ in Eqs.~(A2),~(A3) and~(A4) of~\cite{hinderer-lackey-lang-read} need to be multiplied by a factor of 2, and the one in Eq.~(A5) of~\cite{hinderer-lackey-lang-read} should read $+(9/64 \eta) X_1^{8} (1+18 X_2 +105 X_2^2) (\lambdabarnew_2)^2  x^{15/2}$.

\section{LIGO III Noise Curve Fitting Formula}
\label{app:LIGOIII}

We create a polynomial fitting formula of the LIGO III noise curve~\cite{LIGO3-noise} as
\be
\label{eq:fit-LIGOIII}
\log_{10} \left( \sqrt{S_n(f)} \right) = \sum_{k=0}^{9} c_k (\log_{10} f)^{k}\,, 
\ee
where the coefficients are given in Table~\ref{table:coeff-LIGOIII}. We compare the fit with the data in Fig.~\ref{fig:noise-LIGOIII}.

\bibliography{master}
\end{document}


\title{Erratum: Multipole Love Relations}

\author{Kent Yagi}
\affiliation{Department of Physics, Montana State University, Bozeman, MT 59717, USA.}
\affiliation{Department of Physics, Princeton University, Princeton, NJ 08544, USA.}
\affiliation{Department of Physics, University of Virginia, Charlottesville, Virginia 22904, USA.}

\date{\today}



\maketitle


In Sec.~VA of~\cite{Yagi:2013sva}, we claimed that the magnetic-type tidal perturbation does not affect the conservative sector of binary dynamics for non-rotating bodies, but this was incorrect. Using~\cite{racine}, one finds the radial acceleration due to the tidally-induced $\ell$th magnetic-type multipole moment as $a_r \propto \bar \sigma_\ell v^2/r^{2\ell+3}$, which gives a $2\ell+2$ post-Newtonian (PN) contribution. Comparing this with the $3\ell$~PN dissipative contribution, one finds that the two contributions enter at the same PN order for $\ell = 2$ while the conservative contribution dominates the dissipative one for $\ell \geq 3$. Such a feature is same as the electric-type tidal perturbation case.

Such a conservative contribution changes how $\bar \sigma_2$ enters in the gravitational waveform phase discussed in Appendix~A2 of~\cite{Yagi:2013sva} as follows. 
First, using~\cite{racine}, one finds the radial acceleration due to the tidally-induced quadrupole moment of the 1st body $S_{ij,1}^T$ as
\ba
\label{eq:accel-sigma21}
a_r^{\bar \sigma_2^1} &=& \frac{m}{r_{12}^2} \left[ 1+8 \epsilon_{jmn} v_m \frac{S_{in,1}^{T} n_{ij}}{m_1 r_{12}^2} \right] \nn \\ 
&=&\frac{m}{r_{12}^2} \left[ 1 +48 \, \bar \sigma_2^1 \, \eta \, X_1^3 \, x^6 \right]\,,
\ea
where we correct the tidally-induced quadrupolar current multipole moment as
\be
\label{eq:ST}
S^T_{ij,1} = 12 \frac{m_2}{r_{12}^3} \sigma_2^1 \epsilon_{k\ell( j} n_{i) k} v_\ell\,,
\ee
using Eq.~(6.10) in~\cite{racine}.
This then yields the Kepler's third law and binding energy as
\ba
r_{12} &=& \left( \frac{m}{\omega^2} \right)^{1/3} \left[ 1+16 \, \bar \sigma_2^1 \, \eta  \, X_1^3 x^6 \right]\,, \\
E &=& - \frac{1}{2} \eta m x \left[ 1  - 88 \bar \sigma_2^1 \, \eta  \, X_1^3 x^6 \right]\,,
\ea
respectively. This binding energy can also be derived from the tidal interaction Lagrangian in~\cite{tiziano}.

The above conservative contribution propagates to the dissipative sector. Taking into account corrections to the mass quadrupole moment of a binary system given by~\cite{vines1}
\be
M^{ab}= \eta m r_{12}^2 n^{\langle ab \rangle} +
 \frac{8}{9} \frac{m_2}{m} \left[ 2 v^{i} \epsilon^{ij\langle a} S^{b \rangle j,T}_{1} -  \epsilon^{ij \langle a}\dot S^{b \rangle j,T}_1 n^i r_{12}   \right]\,, \nn
\ee
that was missing in~\cite{Yagi:2013sva}, the luminosity of gravitational waves given in Eq.~(A11) in~\cite{Yagi:2013sva} is corrected to 
\be
\dot{E} = - \frac{32}{5} \eta^2 x^{5} \left[ 1 -\frac{2}{3}  \bar \sigma_2^1\, X_1^4  (X_1-113X_2)  x^6 \right]\,.
\ee
With these expressions at hand, we finally arrive at the gravitational waveform phase including the contribution of $\bar \sigma_2^1$ and Eq.~(A12) in~\cite{Yagi:2013sva} is corrected to
\be
\Psi (f) = \frac{3}{128} \frac{1}{\eta} x^{-5/2} \left[ 1 + \frac{20}{21} \bar \sigma_{2}^1 \, X_1^4 (X_1-1037 X_2)  x^6 \right]\,.
\ee

\emph{Acknowledgments}: We deeply thank Tiziano Abdelsalhin, Paolo Pani and Leonardo Gualtieri for pointing out the mistakes and helping us derive correct expressions. 



\bibliography{master}